\documentclass[12pt]{iopart}
\usepackage{slashed}
\usepackage[colorlinks=true,citecolor=blue,linkcolor=blue]{hyperref}
\usepackage{cite}
\usepackage{url}
\usepackage{multirow}% for tables
\usepackage{graphicx}
\usepackage[utf8]{inputenc}
\usepackage{iopams}  
\begin{document}
\newcommand\redsout{\bgroup\markoverwith{\textcolor{red}{\rule[0.5ex]{4pt}{1pt}}}\ULon}
\title[]{Analysis of $e^{+}e^{-}\rightarrow H^{\pm}W^{\mp}$ in the presence of a circularly polarized laser field}

\author{J. Ou aali,$^1$ M. Ouhammou,$^1$ M. Ouali,$^1$ L. Rahili,$^2$  S. Taj,$^1$ and B. Manaut$^{1,}$}

\address{$^1$ Polydisciplinary Faculty, Laboratory of Research in Physics and Engineering Sciences, Team of Modern and Applied Physics, Sultan Moulay Slimane University, Beni Mellal, 23000, Morocco}
\address{$^2$ EPTHE, Faculty of Sciences, Ibn Zohr University, B.P 8106, Agadir, Morocco.}
\ead{b.manaut@usms.ma}
\vspace{10pt}
%\begin{indented}
%\item[]August 2017
%\end{indented}

\begin{abstract}
In this study, we have investigated the production of a charged Higgs boson in assiciation with a charged weak $W$-boson ($H^{\pm}W^{\mp}$) at tree level via $e^{+}e^{-}$ annihilation in the presence of a laser field with circular polarization. At first step and by using analytical techniques, we have derived the expression of the differential cross section involving both $Z$ and $\gamma$ diagrams. Next, we have shown that the cross section of this process inside the laser field is proportional to the form factor $F_{HWV}$, and it depends on the mass of the charged Higgs boson and the energy of the center of mass. Then, we have analyzed the dependence of the total laser-assisted cross section on the number of photons exchanged for two different cases of form factor. Finally, we have found that the maximum number of photons that can be transferred depends only on the laser parameters.
\end{abstract}
%
%%
%% Uncomment for keywords
\vspace{2pc}

% Uncomment for Submitted to journal title message
%\submitto{\LPL}

 %Uncomment if a separate title page is required
%\maketitle
% 
% For two-column output uncomment the next line and choose [10pt] rather than [12pt] in the \documentclass declaration
%\ioptwocol
%

\section{Introduction}
Since 1960s, the laser technology was and remains the focus of scientists as it receives a great attention among them. Its importance lies not only in studing the physical phenomena associated with laser, but also in assisting our understanding of fundamental particle physics and atomic physics. The first careful studies of laser-matter interaction in non relativistic atomic physics was hypothesized by Francken and Joachain \cite{1}. After, for a well comprehension of the behavior of particles and their properties, some phyisicists have been elaborated laser-atom interaction in relativistic regime. They first studied laser field effects on relativistic atomic physics scattering processes \cite{2,3}. Then, they studied some laser-assisted weak decay processes \cite{4,5}. Due to the significant advancements achieved by laser technology lately\cite{6}, it becomes possible to encourage the laser-assisted production of heavy elementary particles such as neutral higgs pair production\cite{7}, charged higgs pair production\cite{8} and Higgsstrahlung production\cite{9}. This is being achieved by relying on high-intensity laser sources which provide the incident beam $e^{+}e^{-}$ with high kinetic energies. In ref \cite{10}, it is found that the laser field has a great impact on the pion's behavior and its properties such as the decay rate and the particle's lifetime. Moreover, it is shown that the laser field decreases the total cross section of heavy elementary particles' production (see \cite{9,ouali} for more details).

The incredible discovery of the last absent particle of the Standard Model (SM) with a mass around $125$ GeV at the Large Hadron Collider (LHC) at CERN\cite{12,13} opens the door to a deeper understanding of the molecular structure of matter and also phenomena related to high energy processes and particles' behaviors. Despite the fact that the observed properties of the new neutral Higgs boson are in a great agreement with the prediction of the SM, there are some questions that SM does not solve such as the existence of dark matter and the origin of neutrino mass. These unexplained problems in the standard model motivate physicists to continue their researches and experiments for new physics, and to gives information about the extended Higgs sector (Charged Higgs $H^{+}$, $H^{-}$)\cite{14,15} and analyse the coverage of resonant di-Higgs production in the four final state \cite{J.Ouaali,Tania}. In addition, it is well known that the insertion of the laser field in particle physics has a great impact on the particles properties and its behavior. Therefore, the study of laser-assisted high energy processes may lead to significant results such as determining the properties of the charged Higgs boson.

The pair production of charged Higgs boson as well as neutral higgs boson from $e^{-}e^{+}$ annihilation are analyzed within many extensions BSM, whether in Inert Higgs Doublet Model (IHDM)\cite{19}, Two Higgs Doublet Model (2HDM)\cite{20}, or in the SM which is extended with Triplet (HTM)\cite{21}.
Moreover, the process which acts as a source of charged Higgs boson $H^{\pm}$ associated with a $W^{\mp}$ boson at $e^{+}e^{-}$ collider is studied \cite{22}, and the signal process proceeds through $s$-channel via virtual $\gamma$ and $Z$-boson exchange. The authors of this work (ref\cite{22}) showed how much the cross section can be ameliorated by quark- and Higgs-loop effects. Furthermore, the coupling $H^{\pm}W^{\mp}Z$ is the most related to the production of charged Higgs, and, in ref\cite{23}, the authors studied the possibility of measuring it by using the recoil method at ILC. In this respect, we have investigated the production of a charged Higgs bosons in association with a charged weak gauge boson inside a circularly polarized electromagnetic field to analyze theoretically and analytically its effect on the production cross section.

This work is outlined as follows: In section \ref{sec:theory}, we start our discussion by providing the theoretical calculation of the total cross section of charged Higgs production in association with W-boson from $e^{+}e^{-}$ annihilation in the presence of a circularly polarized laser field. Section \ref{sec:results} is devoted to analyze and discuss the obtained results. The conclusion will be presented in section \ref{sec:Conclusion}. It should be pointed out that, in this paper, we have used natural units $\hbar=c=1$, and the metric tensor $g^{\mu\nu}$ is chosen such as $g^{\mu\nu}=(1,-1,-1,-1)$. 

\section{Outline of the theory}\label{sec:theory}
We consider the process which acts as a source of the charged Higgs boson production in association with a $W$ boson at $e^{-}e^{+}$ colliders such as:
\begin{equation}
e^{-}(p_{1}, s_{1})+e^{+}(p_{2}, s_{2})\rightarrow W^{\mp}(k_{1}, \lambda) + H^{\pm}(k_{2}) 
\label{eq1}
\end{equation}
where $p_1$ and $p_2$ are the free four-momentum of incident electron and positron. $k_1$ and $k_2$ indicate the free four-momentum of $W$ boson and the charged Higgs, respectively. Obviously, the charged Higgs boson associated with a $W$ boson are produced via the s-channel process mediated by virtual $\gamma$ and $Z$-boson exchange. These processes are described by the Feynman diagrams which are depicted in figure \ref{fig1}.
\begin{figure}[hptp]
  \centering
      \includegraphics[scale=0.3]{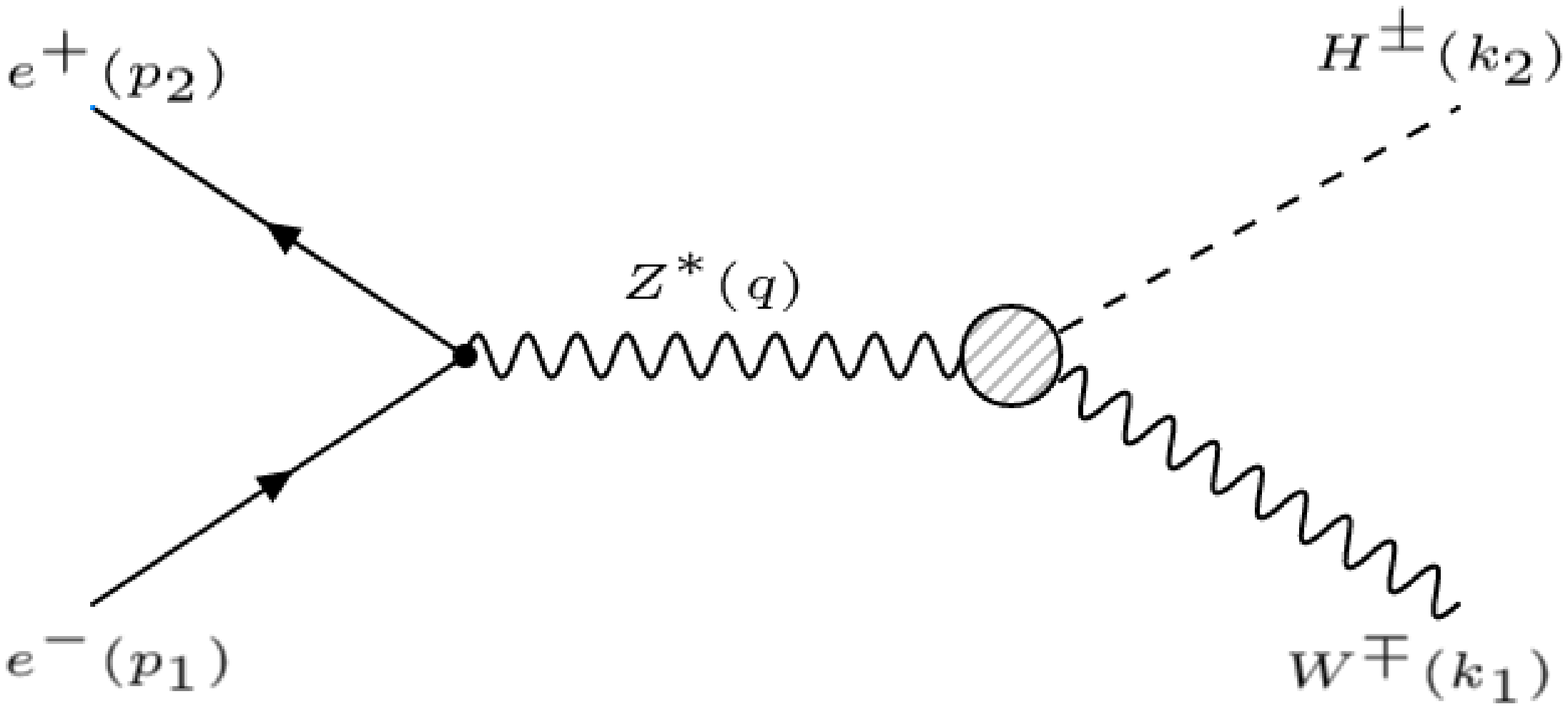}\hspace*{0.4cm}
      \includegraphics[scale=0.3]{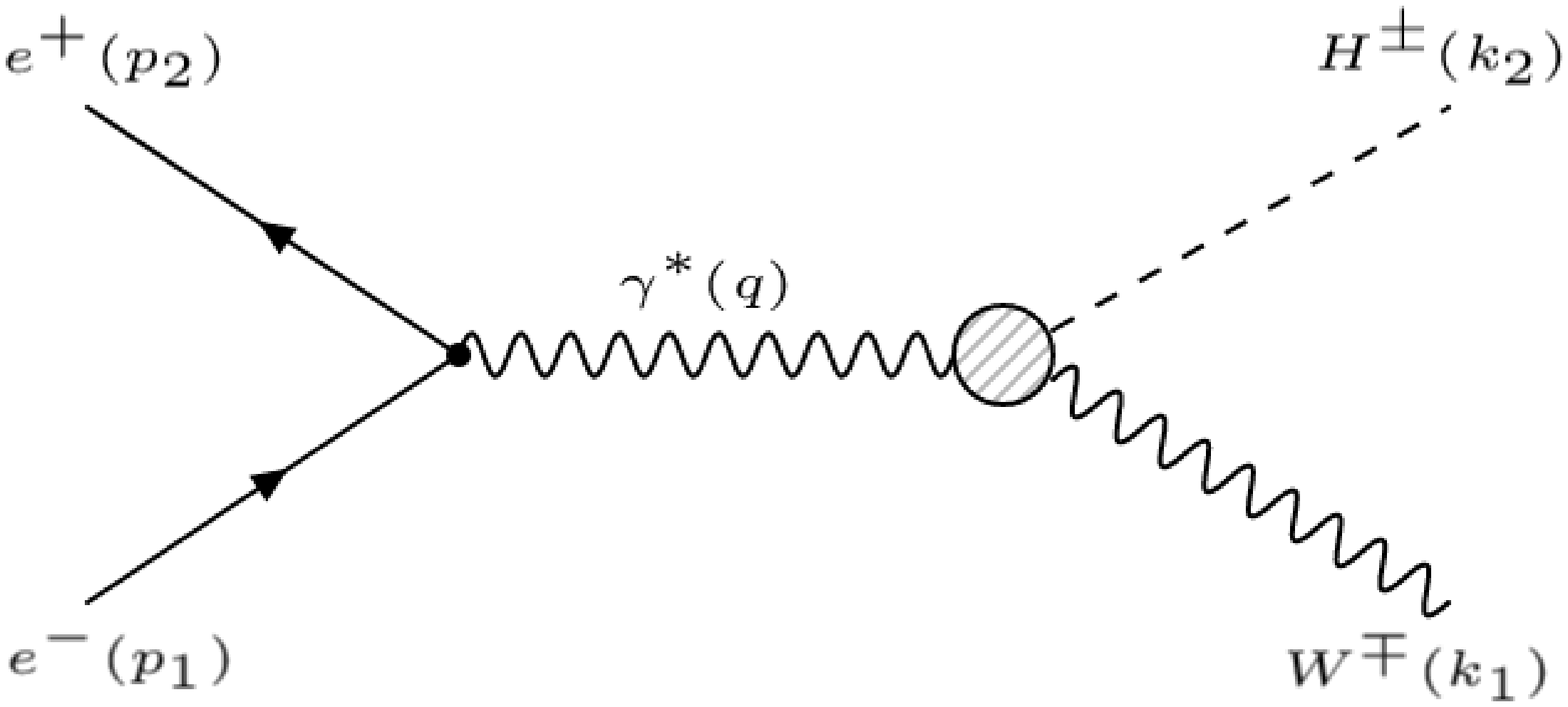}\par\vspace*{0.5cm}
        \caption{Leading-order Feynman diagrams for the electron-positron annihilation process $e^{-}e^{+}\rightarrow H^{\pm}W^{\mp}$.}
        \label{fig1}
\end{figure}

The effective Lagrangian needed in our study \cite{eff1,eff2} is given by the following expression:
\begin{equation}
\fl \eqalign{
\hspace*{2.5cm} \mathcal{L}_{eff}&= gM_{W}f_{HWV}H^{\pm}W^{\mp}_{\mu}V^{\mu}+g_{HWV}H^{\pm}F^{\mu\nu}_{V}F_{W\mu\nu} \cr
&+ (ih_{HWV}\epsilon_{\mu\nu\rho\sigma}F^{\mu\nu}_{V}F^{\rho\sigma}_{W}+h.c),
}
 \label{eff}
\end{equation}
Where, $V = (Z, \gamma)$. $F^{\mu\nu}_{V}$ and $F^{\mu\nu}_{W}$ stand for the field strength tensors for the weak gauge bosons ($Z$ and $W$). Let us remind that the antisymmetric tensors $\epsilon_{\mu\nu\rho\sigma}$ is defined in such a way to satisfy $\epsilon_{0123} = 1$. It should be noted that $f_{HWV}$ is the coefficient of the dimension three operator, while $g_{HWV}$ and $h_{HWV}$ are those of the dimension five operators. Therefore, only the form factor $f_{HWV}$ may appear at the tree level. Besides all mentioned before, the electron and positron are embedded in a circularly polarized laser field which is chosen to be along the $z$-axis. Its polarization vector is given as $k = \omega(1,0,0,1$) where $k^2=0$, and it is described by the following classical four potential:
\begin{equation}
A^{\mu}(\phi) = \chi_{1}^{\mu}\cos\phi + \chi_{2}^{\mu}\sin\phi.
\label{eq4}
\end{equation}
Here, $\phi=k.x$ stands for the phase of laser field, and $\chi_{1,2}^{\mu}$ are the polarization four vectors chosen as $\chi_{1}^{\mu}=(0,\chi,0,0)$ and $\chi_{2}^{\mu}$ = $(0,0,\chi,0)$, with $\chi$ denotes the amplitude of the vector potential.  We assume that $\chi_{1,2}^{\mu}$ are taken in such a manner that the following equations $\chi_{1}^{2}=\chi_{2}^{2}$ = $\chi^{2}$ = $-|\pmb{\chi}|^2$ = $-(\varepsilon_{0}/\omega)^2$ are checked, with $\varepsilon_{0}$ is the electric field's amplitude and $\omega$ its frequency. The application of Lorentz gauge condition $\partial_{\mu}A^{\mu}$ = $0$ implies that $(k.\chi_{1})=(k.\chi_{2})=0$. The scattering-matrix element at tree level \cite{28} for the process represented by figure \ref{fig1} can be expressed as follows:
\small
\begin{equation}
\fl \eqalign{
S_{fi}({e}^{+}{e}^{-}\rightarrow H^{\pm}W^{\mp})=&  \int_{}^{} d^4x  \int {}^{} d^4y \Big\lbrace \bar{\psi}_{p_{2}, s_{2}}(x)   \Big( \frac{-ie}{2C_{W} S_{W} } \gamma^{\mu} (g_v^{e} -g_a^{e}\gamma^{5})   \Big)   \psi_{p_{1}, s_{1}}(x)\cr&\times D_{\mu\nu}(x-y) \varphi^{*\nu}_{k_{1}, \lambda}(y)(igM_{W}f_{HWZ}^{} )\varphi^{*}_{k_2}(y)+\bar{\psi}_{p_{2}, s_{2}}(x)(-ie \gamma^{\mu})\psi_{p_{1}, s_{1}}(x)\cr&\times G_{\mu\nu}(x-y)\varphi^{*\nu}_{k_{1}, \lambda}(y) (igM_{W}f_{HW\gamma}^{})  \varphi^{*}_{k_2}(y)
}
\label{eq5}
\end{equation}
\normalsize
In the equation above, the short notations $C_W$ and $S_W$ stands respectively for $\cos_{\theta_W}$ and $\sin_{\theta_W}$, where $\theta_W$ is the Weinberg angle. $x$ denotes the space time coordinates of the electron and positron, while $y$ is the space-time coordinates of the Higgs bosons and $W$ boson. $g_{v}^{e} = -1 + 4\sin^{2}_{\theta_W}$ and $g_{a}^{e} = 1$ are the vector and axial vector coupling constants. $g$ is the electroweak coupling which is defined by $g^2$ = $e^{2}/\sin^{2}_{\theta_W}=8G_{F}M^{2}_{Z}\cos^{2}_{\theta_W}/\sqrt{2}$.
In the Feynman gauge, the $Z^{*}$-boson propagator $D_{\mu\nu}(x - y)$ and the $\gamma^{*}$-boson propagator $G_{\mu\nu}(x - y)$ are given by \cite{28}:
\begin{equation}
D_{\mu\nu}(x-y)=\int \frac{d^{4}q}{(2\pi)^4} \frac{-ig_{\mu\nu}}{q^{2}-M_{Z}^{2}}e^{-iq(x-y)}
\label{eq6}
\end{equation}
\begin{equation}
G_{\mu\nu}(x-y)=\int \frac{d^{4}q}{(2\pi)^4} \frac{-ig_{\mu\nu}}{q^{2}}e^{-iq(x-y)}
\label{eq7}
\end{equation}
Here, $q$ is the four-momentum of the off-shell $V$ ($V = Z$ or $\gamma$) boson propagators. The quantity $\psi_{p_{1}, s_{1}}(x)$ is the Dirac-Volkov state \cite{28} of the electron, while $\psi_{p_{2}, s_{2}}(x)$ is the Dirac-Volkov state of the positron in the presence of the laser field. Their expressions can be written as:
\begin{equation}
\cases{\psi_{p_{1}, s_{1}}(x)= \Big[1-\frac{e \slashed k \slashed A}{2(k.p_{1})}\Big] \frac{u(p_{1}, s_{1})}{\sqrt{2Q_{1}V}} \exp^{iS(q_{1},s_{1})}&\\
\psi_{p_{2}, s_{2}}(x)= \Big[1+\frac{e \slashed k \slashed A}{2(k.p_{2})}\Big] \frac{v(p_{2}, s_{2})}{\sqrt{2Q_{2}V}} \exp^{iS(q_{2},s_{2})}&\\}
\label{eq8}
\end{equation}
with, $u(p_{1}, s_{1})$ and $v(p_{2}, s_{2})$ are the Dirac spionrs of electron and positron, respectively. $s_{i}$($i$ = $1,2$) are their spins. $p_{1}$ = ($E_{1}$,$|p_{1}|$,0,0) and $p_2$ = ($E_2$,$-|p_{1}|$,0,0) are referring to their corresponding free four momenta in the centre of mass frame. $Q_{1}$ and $Q_{2}$ denote the effective energy acquired by the electron and positron inside the electromagnetic field. We define the arguments of the exponential terms in equation \ref{eq8} as:
\begin{equation}
\cases{S(q_{1},s_{1})=- q_{1}x +\frac{e(\chi_{1}.p_{1})}{k.p_{1}}\sin\phi - \frac{e(\chi_{2}.p_{1})}{k.p_{1}}\cos\phi&\\
S(q_{2},s_{2})=+ q_{2}x +\frac{e(\chi_{1}.p_{2})}{k.p_{2}}\sin\phi - \frac{e(\chi_{2}.p_{2})}{k.p_{2}}\cos\phi&\\}
\label{eq9}
\end{equation}
where, $q_i$, $i$ = (1,2) stands respectively for the effective momentum of electron and positron inside the laser field.  They are related to their corresponding free four momentum such that:
\begin{equation}
q_{i} = p_{i} + \frac{e^{2}a^2}{2(k.p_{i})}k.
\label{eq10} 
\end{equation}
In this case, the square of the effective momentum reads as, $q_{i}^{2}=m_{e}^{*2}$ = ($m_{e}^{2} + e^{2}a^{2}$), with $m_{e}^{*}$ is the effective mass of incident particles, $m_e$ is the electron mass and $e$ denotes its charge. Moreover, in our study, we consider that there is no interaction between the produced particles and the laser field. Therefore, they are described by Klein-Gordon states such that:
\begin{equation}
\varphi^{\mu}_{k_{1}, \lambda}(y)=\frac{\epsilon^{\mu}(k_{1}, \lambda)}{\sqrt{2 E_{W^{\mp}} V}} e^{-ik_{1}y} \hspace*{0.1cm},\hspace*{0.5cm}  \\\ \varphi_{k_{2}}(y)=\frac{1}{\sqrt{2 E_{H^{\pm}} V}} e^{-ik_{2}y},
\label{eq11}
\end{equation}
with, $k_{1}$ = ($E_{W^{\mp}}$,$|k_1|\cos\theta$,$|k|\sin\theta$,0) and $k_2$ = ($E_{H^{\pm}}$,$-|k_2|\cos\theta$,$-|k_2|\sin\theta$,0) are respectively the free four-momentum of $W$-boson and the charged Higgs-boson. $\epsilon^{\mu}(k_1, \lambda)$ stands for $W$-boson polarization vector where $\lambda$ denotes its polarization such that \small{$\sum_{\lambda=1}^{3}\epsilon^{\mu}(k_{1}, \lambda)\epsilon^{*\nu}(k_{1}, \lambda)=-g^{\mu\nu}+k_{1}^{\mu}k_{1}^{\nu}/M_{W}^{2}$}. Then, to determine the expression of the scattering matrix element, we will insert the equations given by \ref{eq6}, \ref{eq7}, \ref{eq8}, \ref{eq9} \ and \ref{eq11} into equation \ref{eq5}, and we get:
\begin{equation}
S_{fi}^{n}({e}^{+}{e}^{-}\rightarrow H^{\pm}W^{\mp})=\frac{(2\pi)^{4}\delta^{4}(k_{1}+k_{2} -q_{1}-q_{2}-nk)}{4V^{2}\sqrt{Q_{1}Q_{2}E_{H^{\pm}}E_{W^{\mp}}}} \big(M_{Z}^{n} + M_{\gamma}^{n} \big)
\label{12}  
\end{equation}
where $n$ indicates the number of exchanged photons between the laser field and the colliding physical
system. The two parts which constitute the total scattering amplitude $M_{Z}^{n}$ and $M_{\gamma}^{n}$ are coming from the contribution of the free-photon exchange and the $Z$-boson contribution, respectively. The first part can be expressed in terms of Bessel functions as follows:
\begin{equation}
\fl\eqalign{
M_{Z}^{n}&= \frac{e}{2C_{W}S_{W}}  gM_{W}F_{HWZ}  \frac{\epsilon^{*\mu}(k_{1},\lambda)}{(q_{1}+q_{2}+nk)^{2}-M_{Z}^{2}}  \Bigg\lbrace\bar{v}(p_{2}, s_{2}) \cr &\times \Bigg[ \kappa^{0}_{\mu}\,J_{n}(z)e^{-in\phi _{0}}(z)+ \frac{1}{2} \,\, \kappa^{1}_{\mu}\Big(J_{n+1}(z)e^{-i(n+1)\phi _{0}} + J_{n-1}(z)e^{-i(n-1)\phi _{0}}\Big) \cr &+ \frac{1}{2\, i}\, \kappa^{2}_{\mu}\Big(J_{n+1}(z)e^{-i(n+1)\phi _{0}}-J_{n-1}(z)e^{-i(n-1)\phi _{0}}\Big)\Bigg] u(p_{1}, s_{1})  \Bigg\rbrace,
}
\label{13}
\end{equation}
where, the quantities $\kappa^{0}_{\mu}$, $\kappa^{1}_{\mu}$ and $\kappa^{2}_{\mu}$ can be written as follows: 
\begin{equation}
\cases{\kappa^{0}_{\mu}=\gamma_{\mu}(g_{v}^{e}-g_{a}^{e}\gamma^{5})+2a_{p_{1}}a_{p_{2}}\chi^{2}k_{\mu}\slashed k(g_{v}^{e}-g_{a}^{e}\gamma^{5})&\\
\kappa^{1}_{\mu}=a_{p_{1}}\gamma_{\mu}(g_{v}^{e}-g_{a}^{e}\gamma^{5})\slashed k\slashed \chi_{1}-a_{p_{2}}\slashed \chi_{1}\slashed k \gamma_{\mu}(g_{v}^{e}-g_{a}^{e}\gamma^{5})&\\
\kappa^{2}_{\mu}=a_{p_{1}}\gamma_{\mu}(g_{v}^{e}-g_{a}^{e}\gamma^{5})\slashed k\slashed \chi_{2}-a_{p_{2}}\slashed \chi_{2}\slashed k \gamma_{\mu}(g_{v}^{e}-g_{a}^{e}\gamma^{5})& \\}
\label{14}
\end{equation}
with $a_{p_{i}}=e/2(kp_{i})$($ i=1,2 $). Similarly, one gets $M_{\gamma}^{n}$ as:
\begin{equation}
\fl\eqalign{
M_{\gamma}^{n}&= \frac{e}{(q_{1}+q_{2}+nk)^{2}} gM_{W}F_{HW\gamma} \epsilon^{*\mu}(k_{1},\lambda) \Bigg\lbrace\bar{v}(p_{2}, s_{2})\Bigg[ \xi^{0}_{\mu}\,J_{n}(z)e^{-in\phi _{0}}(z) \cr&+ \frac{1}{2} \,\, \xi^{1}_{\mu}\Big(J_{n+1}(z)e^{-i(n+1)\phi _{0}} + J_{n-1}(z)e^{-i(n-1)\phi _{0}}\Big)\nonumber + \frac{1}{2\, i}\,\xi^{2}_{\mu}\cr &\times \Big(J_{n+1}(z)e^{-i(n+1)\phi _{0}}-J_{n-1}(z)e^{-i(n-1)\phi _{0}}\Big)\Bigg] u(p_{1}, s_{1}) \Bigg\rbrace.
}
\label{15}
\end{equation}
In equation \ref{15}, the factors $\xi^{0}_{\mu}$, $\xi^{1}_{\mu}$ and $\xi^{2}_{\mu}$ are related to Dirac matrices as follows:
\begin{equation}
\cases{\xi^{0}_{\mu}=\gamma_{\mu}+2a_{p_{1}}a_{p_{2}}\chi^{2}k_{\mu}\slashed k &\\
\xi^{1}_{\mu}=a_{p_{1}}\gamma_{\mu}\slashed k\slashed \chi_{1}-a_{p_{2}}\slashed \chi_{1}\slashed k \gamma_{\mu}&\\
\xi^{2}_{\mu}=a_{p_{1}}\gamma_{\mu}\slashed k\slashed \chi_{2}-a_{p_{2}}\slashed \chi_{2}\slashed k \gamma_{\mu}& \\}
\label{16}
\end{equation}
The two form factors $F_{HWZ}$ and $F_{HW\gamma}$ in equations \ref{13} and \ref{15} are successively related to the coefficients $f_{HWZ}$ and $f_{HW\gamma}$ that appear in the effective Lagrangian given by the equation \ref{eff}. Moreover, the reason of $U(1)_{em}$ gauge invariance imposes that the form factor $F_{HW\gamma}$ should be vanished in any extended model beyond the SM at the tree level. So, the $Z$-boson couplings has a dominant contribution. However, due to the violation of the Custiodal Symmetry in the kinetic term, all the form factors containing $F_{HWZ}$ are zero at the tree level in some extended Higgs models with only two doublet scalar fields such as two-Higgs-doublet model (2HDM) and beyond. The simplest models in which the form factor $F_{HWZ}$ appears at the tree level are given in \cite{23,25}.
The argument of the Bessel function $z$ and the phase $\phi_{0}$ are determined by, $ z=\sqrt{\alpha_{1}^{2}+\alpha_{2}^{2}}$ and $\phi_{0}= \arctan(\alpha_{2}/\alpha_{1})$, where:
\begin{equation}
\alpha_{1}=\frac{e(\chi_{1}.p_{1})}{(k.p_{1})}-\frac{e(\chi_{1}.p_{2})}{(k.p_{2})} \hspace*{0.1cm},\hspace*{0.5cm}  \\\ \alpha_{2}=\frac{e(\chi_{2}.p_{1})}{(k.p_{1})}-\frac{e(\chi_{2}.p_{2})}{(k.p_{2})},
\label{alpha}
\end{equation}
In general, the differential cross section in the center of mass frame is calculated by dividing the squared matrix element of equation \ref{12} by $V T$, and by the current of incident particles defined by $|J_{inc}|=(\sqrt{(q_{1}q_{2})^{2}-m_{e}^{*^{4}}}/{Q_{1}Q_{2}V})$, then by the density of the particles $\rho = V^{-1}$. We get the expression of the partial differential cross section as follows:
\begin{equation}
d\sigma_{n}=\frac{|S_{fi}^{n}|^{2}}{VT}\frac{1}{|J_{inc}|}\frac{1}{\varrho}V\int_{}\frac{d^{3}k_{2}}{(2\pi)^3}V\int_{}\frac{d^{3}k_{1}}{(2\pi)^3}.
\label{17}
\end{equation} 
After some simplification and by summing over the final states and averaging over the initial ones, the differential cross section becomes:
\begin{equation}
\fl\eqalign{ 
d\bar{\sigma_{n}}({e}^{+}{e}^{-}\rightarrow H^{\pm}W^{\mp})&=\frac{1}{16\sqrt{(q_{1}q_{2})^2-m_{e}^{*^{4}}}}   \big|\overline{M_{Z}^{n} + M_{\gamma}^{n}} \big|^{2} \int_{}\frac{|\mathbf{k}_{2}|^{2}d|\mathbf{k}_{2}|d\Omega}{(2\pi)^2E_{H^{\pm}}}\int_{}\frac{d^{3}k_{1}}{ E_{W^{\mp}}} \cr  &\times \delta^{4}(k_{1}+k_{2}-q_{1}-q_{2}-nk).
}
\label{18}
\end{equation}
The remaining integral over $d^{3}k_{1}$ can be estimated by using the well known formula \cite{28} given by:
\normalsize
\begin{equation}
 \int d\mathbf y f(\mathbf y) \delta(g(\mathbf y))=\frac{f(\mathbf y)}{|g^{'}(\mathbf y)|_{g(\mathbf y)=0}}.
 \label{19}
\end{equation}
Finally, we find the final expression of the differential cross section as follows:
\small
\begin{equation}
\fl\eqalign{
\frac{d\sigma_{n}}{d\Omega}({e}^{+}{e}^{-}\rightarrow H^{\pm}W^{\mp})&=\frac{1}{16\sqrt{(q_{-}q_{+})^2-m_{e}^{*^{4}}}}   \big|\overline{M_{Z}^{n} + M_{\gamma}^{n}} \big|^{2} \frac{2|\mathbf{k}_{2}|^{2}}{(2\pi)^2E_{H^{\pm}}}\frac{1}{|g^{'}(|\mathbf{k}_{2}|)|_{g(|\mathbf{k}_{2}|)=0}},
}
\label{20}
\end{equation}
\normalsize
where the function $g^{'}(|\mathbf{k}_{2}|)$ is given by:
\begin{equation}
\big|g^{'}(|\textbf{k}_{2}|)\big|=-4\Big[\frac{(\sqrt{s}+n\omega)|\textbf k_{2}|}{2\sqrt{|\textbf k_{2}|^{2}+M_{H^{\pm}}^{2}}}-\frac{|\textbf k_{2}|}{\sqrt{|\textbf k_{2}|^{2}+M_{H^{\pm}}^{2}}}\frac{e^{2}a^{2}}{\sqrt{s}}\Big].
\label{23}
\end{equation}
By using FeynCalc tools \cite{Feyncalc}, we have derived the expression of the quantity $\big|\overline{M_{Z}^{n} + M_{\gamma}^{n}} \big|^{2}$ given in equation \ref{20}, and it can be written as:
\small
\begin{equation}
\fl\eqalign{
 \big|\overline{M_{Z}^{n} + M_{\gamma}^{n}} \big|^{2}&=\nonumber\frac{1}{4}\sum_{n=-\infty}^{+\infty}\sum_{s}\big| M_{Z}^{n} + M_{\gamma}^{n} \big|^{2}=\frac{1}{4}\sum_{n=-\infty}^{+\infty}\Bigg\lbrace(2gM_{W})^{2}(F_{HW\gamma})^{2}\frac{e^{4}}{(q_{1}+q_{2}+nk)^{4}}\cr &\times  Tr\Bigg[(\slashed p_{1}-m_{e})\Big[ \xi^{0}_{\mu}\,J_{n}(z)e^{-in\phi _{0}}(z)+\xi^{1}_{\mu}\,\,\frac{1}{2}\Big(J_{n+1}(z)e^{-i(n+1)\phi _{0}} + J_{n-1}(z)e^{-i(n-1)\phi _{0}}\Big) \cr &+ \xi^{2}_{\mu}\,\frac{1}{2\, i}\Big(J_{n+1}(z)e^{-i(n+1)\phi _{0}}-J_{n-1}(z)e^{-i(n-1)\phi _{0}}\Big)\Big](\slashed p_{2}+m_{e})  \Big[ \xi^{0}_{\nu}\,J^{*}_{n}(z)e^{+in\phi _{0}}(z)\cr &+ \xi^{1}_{\nu}\,\,\frac{1}{2}\Big(J^{*}_{n+1}(z)e^{+i(n+1)\phi _{0}} + J^{*}_{n-1}(z)e^{+i(n-1)\phi _{0}}\Big)-\xi^{2}_{\nu}\,\frac{1}{2\, i}\Big(J^{*}_{n+1}(z)e^{+i(n+1)\phi _{0}}\cr &- J^{*}_{n-1}(z)e^{+i(n-1)\phi _{0}}\Big)\Big]\Bigg]\Big(-g^{\mu\nu}+\frac{(k_{1}^{\mu}.k_{1}^{\nu})}{M_{W}^{2}}\Big)
+ (2gM_{W})^{2}(F_{HWZ})^{2}\left(\frac{e}{2C_{W}S_{W}}\right)^{2}\cr &\times  \left(\frac{1}{(q_{1}+q_{2}+nk)^{2}-M_{Z}^{2}}\right)^{2} Tr\Bigg[(\slashed p_{1}-m_{e})\Big[ \kappa^{0}_{\mu}\,J_{n}(z)e^{-in\phi _{0}}(z)+\kappa^{1}_{\mu}\,\,\frac{1}{2}\Big(J_{n+1}(z)e^{-i(n+1)\phi _{0}} \cr &+ J_{n-1}(z)e^{-i(n-1)\phi _{0}}\Big)+\kappa^{2}_{\mu}\,\frac{1}{2\, i}\Big(J_{n+1}(z)e^{-i(n+1)\phi _{0}}-J_{n-1}(z)e^{-i(n-1)\phi _{0}}\Big)\Big]\cr &\times (\slashed p_{2}+m_{e}) \Big[ \kappa^{0}_{\nu}\,J^{*}_{n}(z)e^{+in\phi _{0}}(z)+ \kappa^{1}_{\nu}\,\,\frac{1}{2}\Big(J^{*}_{n+1}(z)e^{+i(n+1)\phi _{0}} + J^{*}_{n-1}(z)e^{+i(n-1)\phi _{0}}\Big) \cr &- \kappa^{2}_{\nu}\,\frac{1}{2\, i}\Big(J^{*}_{n+1}(z)e^{+i(n+1)\phi _{0}}-J^{*}_{n-1}(z)e^{+i(n-1)\phi _{0}}\Big)\Big]\Bigg]\Big(-g^{\mu\nu}+\frac{(k_{1}^{\mu}.k_{1}^{\nu})}{M_{W}^{2}}\Big)
\cr &+ (2gM_{W})^{2}(F_{HWZ}F_{HW\gamma})\frac{e^{2}}{(q_{1}+q_{2}+nk)^{2}}\left(\frac{e}{2C_{W}S_{W}}\right)   \frac{1}{(q_{1}+q_{2}+nk)^{2}-M_{Z}^{2}}\cr &\times Tr\Bigg[(\slashed p_{1}-m_{e})\Big[ \xi^{0}_{\mu}\,J_{n}(z)e^{-in\phi _{0}}(z) +\xi^{1}_{\mu}\,\,\frac{1}{2}\Big(J_{n+1}(z)e^{-i(n+1)\phi _{0}} + J_{n-1}(z)e^{-i(n-1)\phi _{0}}\Big)\cr &+ \xi^{2}_{\mu}\,\frac{1}{2\, i}\Big(J_{n+1}(z)e^{-i(n+1)\phi _{0}}-J_{n-1}(z)e^{-i(n-1)\phi _{0}}\Big)\Big](\slashed p_{2}+m_{e}) \Big[ \kappa^{0}_{\nu}\,J^{*}_{n}(z)e^{+in\phi _{0}}(z)\cr &+ \kappa^{1}_{\nu}\,\,\frac{1}{2}\Big(J^{*}_{n+1}(z)e^{+i(n+1)\phi _{0}} + J^{*}_{n-1}(z)e^{+i(n-1)\phi _{0}}\Big)- \kappa^{2}_{\nu}\,\frac{1}{2\, i}\Big(J^{*}_{n+1}(z)e^{+i(n+1)\phi _{0}}\cr &- J^{*}_{n-1}(z)e^{+i(n-1)\phi _{0}}\Big)\Big]\Bigg]\Big(-g^{\mu\nu}+\frac{(k_{1}^{\mu}.k_{1}^{\nu})}{M_{W}^{2}}\Big)
+ (2gM_{W})^{2}(F_{HW\gamma}F_{HWZ})\frac{e^{2}}{(q_{1}+q_{2}+nk)^{2}}\cr &\times \left(\frac{e}{2C_{W}S_{W}}\right)  \frac{1}{(q_{1}+q_{2}+nk)^{2}-M_{Z}^{2}} Tr\Bigg[(\slashed p_{1}-m_{e}) \Big[ \kappa^{0}_{\mu}\,J_{n}(z)e^{-in\phi _{0}}(z)\cr &+ \kappa^{1}_{\mu}\,\,\frac{1}{2}\Big(J_{n+1}(z)e^{-i(n+1)\phi _{0}}+ J_{n-1}(z)e^{-i(n-1)\phi _{0}}\Big)+ \kappa^{2}_{\mu}\,\frac{1}{2\, i}\Big(J_{n+1}(z)e^{-i(n+1)\phi _{0}}\cr &- J_{n-1}(z)e^{-i(n-1)\phi _{0}}\Big)\Big] (\slashed p_{2}+m_{e}) \Big[ \xi^{0}_{\nu}\,J^{*}_{n}(z)e^{+in\phi _{0}}(z)+\xi^{1}_{\nu}\,\,\frac{1}{2}\Big(J^{*}_{n+1}(z)e^{+i(n+1)\phi _{0}} \cr &-  J^{*}_{n-1}(z)e^{+i(n-1)\phi _{0}}\Big) -\xi^{2}_{\nu}\,\frac{1}{2\, i}\Big(J^{*}_{n+1}(z)e^{+i(n+1)\phi _{0}}-J^{*}_{n-1}(z)e^{+i(n-1)\phi _{0}}\Big)\Big]\Bigg]\Big(-g^{\mu\nu}+\frac{(k_{1}^{\mu}.k_{1}^{\nu})}{M_{W}^{2}}\Big) \Bigg\rbrace.
}
\end{equation}
\normalsize 
\section{Results and discussion}\label{sec:results}
In this section, we have performed a systematic numerical scan to investigate the scenario which can lead to significant cross section for the production process of the charged Higgs boson associated with a $W$-boson from $e^{+}e^{-}$ annihilation. As we have shown in equation \ref{20}, in the centre of mass frame, this cross section depends on the mass of charged Higgs boson $M_{H^{\pm}}$, the centre of mass energy $\sqrt{s}$ and the laser field parameters such as the number of exchanged photons $n$, the laser field strength $\varepsilon_{0}$ and its frequency $\omega$. Moreover, we can get the total cross section by numerically integrating the equation \ref{20} over the solid angle defined by $d\Omega=\sin(\theta)d\theta d\phi$, where $\theta$ is the angle between the momentum of the outgoing Higgs-bosons and the beam axis. The SM parameters used in our analysis such as $m_{e}=0.511\, MeV$, $M_{Z}=91.1875\, GeV$, $M_{W}=80.379\, GeV$ and $\theta_W=28.75^{\circ}$ are taken from PDG \cite{29}. As it is previously mentioned, we assume that $g_{HWV}=h_{HWV}=0$. Consequenctly, the cross section is proportional to $F_{HWV}$, $V$ = ($Z$ or $\gamma$). Throughout this work, we have considered two sample benchmark points, namely $F_{HW\gamma}$ = $1$ and $F_{HW\gamma}=0$, while $F_{HWZ}=1$\cite{23}.
\begin{table}[hptp]
\centering
\caption{\label{tab1}Laser-assisted total cross section as a function of the number of exchanged photons for different laser field strengths and frequencies by taking $F_{HW\gamma}=1$. The centre of mass energy and the charged Higgs mass are chosen as: $\sqrt{s}=300\, GeV$ ; $M_{H^{\pm}}=200\, GeV$.}
\begin{tabular}{ccccccccc}
\hline
  & & $ \sigma $[fb]  & &$ \sigma $[fb] & & $ \sigma $[fb] &  \\
 $ \varepsilon_{0}(\,V.cm^{-1}) $ &&  He:Ne Laser & & Nd:YAG laser & &  $ CO_{2} $ Laser &\\
& n &  $\omega=2\,eV $ & n &$\omega=1.17\,eV $& n & $\omega=0.117\,eV  $ & \\
 \hline
 \hline
 \hline
 &$\pm7$ & $ 0 $ & $\pm20$ & $ 0 $ & $\pm1200$ & $ 0 $ \\
 &$\pm6$ & $ 0 $ & $\pm15$ & $ 0 $ & $\pm1100$ & $ 0 $ \\
   &$\pm5$ & $ 1.44717 $ & $\pm12$ & $ 1.24094 $ & $\pm900$ & $ 0.00167942 $ \\
  $ 10^{5} $  &$\pm4$ & $ 9.36089 $ & $\pm9$ & $ 20.568 $ & $\pm600$ & $ 0.165453 $ \\
   &$\pm3$ & $ 33.5847 $ & $\pm6$ & $ 0.893134 $ & $\pm300$ & $ 0.104116 $ \\
    &$\pm2$ & $ 47.3638 $ & $\pm3$ & $ 2.63859 $ & $\pm100$ & $ 0.0191935 $ \\
    &$0$ & $ 32.4356  $ & $0$ & $ 13.9913 $  & $0$ & $ 0.134855 $ \\
     \hline
     &$\pm40$ & $ 0 $ & $\pm120$ & $ 0 $ & $\pm6000$ & $ 0 $ \\
 &$\pm35$ & $ 0 $ & $\pm110$ & $ 0 $ & $\pm5100$ & $ 0 $ \\
   &$\pm32$ & $ 9.62478 $ & $\pm80$ & $ 1.98117 $ & $\pm4000$ & $ 0.00227526 $ \\
  $ 10^{6} $  &$\pm24$ & $ 0.782794 $ & $\pm60$ & $ 1.6538 $ & $\pm3000$ & $ 0.0141595 $ \\
   &$\pm16$ & $ 4.39571 $ & $\pm40$ & $ 1.02198 $ & $\pm2000$ & $ 0.00252853 $ \\
   &$\pm8$ & $ 3.06533 $ & $\pm20$ & $ 1.30663 $ & $\pm1000$ & $ 0.000287841 $ \\
    &$0$ & $ 3.5543  $ & $0$ & $ 0.565689 $  & $0$ & $ 0.0117641 $ \\
     \hline
\end{tabular}
\end{table}
\begin{table}[hptp]
 \centering
\caption{\label{tab2}Laser-assisted total cross section as a function of the number of exchanged photons for different laser field strengths and frequencies by taking $F_{HW\gamma}=0$. The centre of mass energy and the charged Higgs mass are chosen as: $\sqrt{s}=300\, GeV$ ; $M_{H^{\pm}}=200\, GeV$.}
\begin{tabular}{ccccccccc}
\hline
  & & $ \sigma $[fb]  & &$ \sigma $[fb] & & $ \sigma $[fb] &  \\
 $ \varepsilon_{0}(\,V.cm^{-1}) $ &&  He:Ne Laser & & Nd:YAG laser & &  $ CO_{2} $ Laser &\\
& n &  $\omega=2\,eV $ & n &$\omega=1.17\,eV $& n & $\omega=0.117\,eV  $ & \\
 \hline
 \hline
 \hline
 &$\pm7$ & $ 0 $ & $\pm20$ & $ 0 $ & $\pm1200$ & $ 0 $ \\
 &$\pm6$ & $ 0 $ & $\pm15$ & $ 0 $ & $\pm1100$ & $ 0 $ \\
   &$\pm5$ & $ 1.19267 $ & $\pm12$ & $ 1.02271 $ & $\pm900$ & $ 0.00138408 $ \\
  $ 10^{5} $  &$\pm4$ & $ 7.71467 $ & $\pm9$ & $ 16.9509 $ & $\pm600$ & $ 0.136356 $ \\
   &$\pm3$ & $ 27.6784 $ & $\pm6$ & $ 0.736066 $ & $\pm300$ & $ 0.0858061 $ \\
    &$\pm2$ & $ 39.0343 $ & $\pm3$ & $ 2.17457 $ & $\pm100$ & $ 0.0158181 $ \\
    &$0$ & $ 26.7314  $ & $0$ & $ 11.5307 $  & $0$ & $ 0.111139 $ \\
     \hline
     &$\pm40$ & $ 0 $ & $\pm120$ & $ 0 $ & $\pm6000$ & $ 0 $ \\
 &$\pm35$ & $ 0 $ & $\pm110$ & $ 0 $ & $\pm5100$ & $ 0 $ \\
   &$\pm32$ & $ 7.93215 $ & $\pm80$ & $ 1.63275 $ & $\pm4000$ & $ 0.00187531 $ \\
  $ 10^{6} $  &$\pm24$ & $ 0.645131 $ & $\pm60$ & $ 1.36296 $ & $\pm3000$ & $ 0.0116694 $ \\
   &$\pm16$ & $ 3.62268 $ & $\pm40$ & $ 0.842252 $ & $\pm2000$ & $ 0.00208386 $ \\
   &$\pm8$ & $ 2.52625 $ & $\pm20$ & $ 1.07684 $ & $\pm1000$ & $ 0.000237221 $ \\
    &$0$ & $ 2.92923 $ & $0$ & $ 0.46206 $  & $0$ & $ 0.0096528 $ \\
     \hline
\end{tabular}
\end{table}

In Tables \ref{tab1} and \ref{tab2}, we present the variation of the partial total cross section of the process $e^{+}e^{-}\rightarrow H^{\pm}W^{\mp}$ as a function of the number of exchanged photons inside the electromagnetic field. This variation is presented for different laser parameters such as the laser field strength and its frequency. Table \ref{tab1} illustrates the case where $F_{HW\gamma}=1$, while Table \ref{tab2} represents the results obtained for $F_{HW\gamma}=0$. As we can see from these tables, the number of photons that can be exchanged varies by changing the laser parameters, and, as a consequence, the partial total cross section changes. Moreover, when we increase the number of exchanged photons, this partial total cross section progressively decreases until it becomes null. This result allows us to define the "cutoff" as the greatest value of photons' number that can be exchanged and for which the partial total cross section becomes null. We note that the cross section takes the same values for the absorption of photons ($+n$) as well as for the emission of photons ($-n$). We remark also that the value of the "cutoff" increases as long as the intensity of the laser field is increased or by decreasing its frequency. Furthermore, by comparing the "cutoffs" given in tables \ref{tab1} and \ref{tab2} for a specified values of $\varepsilon_0$ and $\omega$, we can see that they do not change. For instance, for $\varepsilon_{0}=10^{6}\, V.cm^{-1}$ and $\omega=2\, eV$, the cutoffs are equal to $\pm35$ for both $F_{HW\gamma}=1$ and $F_{HW\gamma}=0$. Therefore, the cutoff does not depend on the choice of $F_{HW\gamma}$, and it depends only on the laser field parameters. To discuss these results more deeply, we have plotted in figure \ref{fig2} the partial total cross section as a function of the laser photons' number for both $F_{HW\gamma}$ = $1$ (green curve) and $F_{HW\gamma}=0$ (blue curve). The laser parameters are chosen as $\varepsilon_{0}=10^{6}\, V.cm^{-1}$ and $\omega=1.17\, eV$.
As we can see from this figure, the maximum number of photons transferred between the laser beam and the colliding system, which cancels the partial total cross section in both cases $F_{HW\gamma}=1$ and $F_{HW\gamma}=0$, is equal to $\pm 110$. Therefore, theses results are in good agreement with those given in tables \ref{tab1} and \ref{tab2}. 
In figure \ref{fig2}, the charged Higgs mass is chosen as $M_{H^{\pm}}=200$ GeV which is consistent with the IHDM constraints \cite{30}.
\begin{figure}[h!]
\centering
      \includegraphics[scale=0.7]{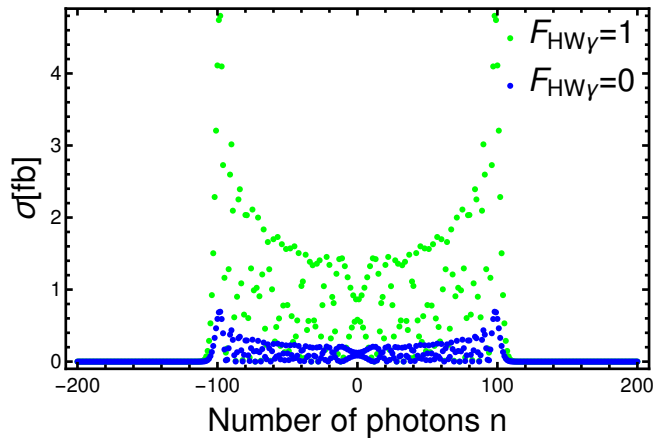}
  \caption{Laser-assisted partial total cross section versus the number of exchanged photons in the case $F_{HWZ}=1$. The green color shows the case $F_{HW\gamma} = 1$, and the blue one shows the case $F_{HW\gamma}= 0 $. The laser field strength and its frequency are chosen as $\varepsilon_0=10^{5}\, V.cm^{-1}$ and $\omega=1.17\, eV$. The centre of mass energy and the charged Higgs mass are taken as $\sqrt{s}=300\, GeV$ and $M_{H^{\pm}}=200\, GeV$.}
  \label{fig2}
\end{figure}
Now, we will discuss the dependence of the laser-assisted total cross section of the process $e^{+}e^{-}\rightarrow H^{\pm}W^{\mp}$ on the centre of mass energy for different transferred photons number and for different laser field strengths and frequencies in the case $F_{HWZ}=1$ and $F_{HW\gamma}=0$. The phenomenology of these scenarios is illustrated in figure \ref{fig3}. Let's start our discussion with the laser-free total cross section. As it is shown in this figure, the cross section increases severely for low centre of mass energies region $\sqrt{s}<326$ GeV until it reaches its maximum at $\sqrt{s}$ $\simeq$ $326$ GeV. Then, it decreases gradually as long as $\sqrt{s}$ increases.
The existence of the peak at $\sqrt{s}$ $\simeq$ $326$ GeV is a consequence of competition among the phase-space enlargement and $s$-channel suppression as much as $\sqrt{s}$ increases\cite{31,32}.
By applying the circularly polarized laser field, we notice that the behavior of the production cross section as a function of $\sqrt{s}$ for different values of the laser's parameters and different exchanged photons number are quite similar. However, compared to its corresponding laser-free cross section, we can see that the laser-assisted cross section decreases by several orders of magnitude. The maximum value of the cross section can surpass $66fb$ for $n=\pm 4$, and it decreases to about $60fb$ and $50fb$ for $n=\pm 3$ and $n=\pm 2$, respectively. As shown in figure \ref{fig3}(a) in which $\varepsilon_0=10^{5}\, V.cm^{-1}$ and $\omega=2\, eV$, the maximal value is reached around $68fb$ for the region of high $n$ ($n=\pm 6$). In this region, the total cross section will be equal to its corresponding laser-free total cross section in all centre of mass energies. This means that the effect of laser field on the incident particles is strongly suppressed. As a result no photon will be exchanged between the laser field and the incident particles. In figures \ref{fig3}(b), \ref{fig3}(c) and \ref{fig3}(d), the total cross section increases as long as the number $n$ raises until $n=\pm$ cutoff similarly to figure \ref{fig3}(a). This summation over $\pm$ cutoff number is called sum-rule \cite{33}, and it leads to a cross section which is equal to the laser-free cross section in all centre of mass energies. Moreover, the total cross section strongly depends on the values of the laser field amplitude and its frequency. By comparing figures \ref{fig3}(a) and \ref{fig3}(b), we remark that, for the same laser field amplitude which is $\varepsilon_0=10^{5}\, V.cm^{-1}$ and for the same number of exchanged photons such as $n=\pm 3$, the corresponding cross section is about $61fb$ for $\omega$ = $2eV$ and $14fb$ for $\omega=1.17\, eV$. 
Another important point to be mentioned, here, is that the peak does not change for all laser field's parameters. This result push us to ask an important question which is: How can these free-parameters change the behavior and properties of the cross section of the studied process?
\begin{figure}[h!]
\centering
  \begin{minipage}[t]{0.478\textwidth}
  \centering
    \includegraphics[width=\textwidth]{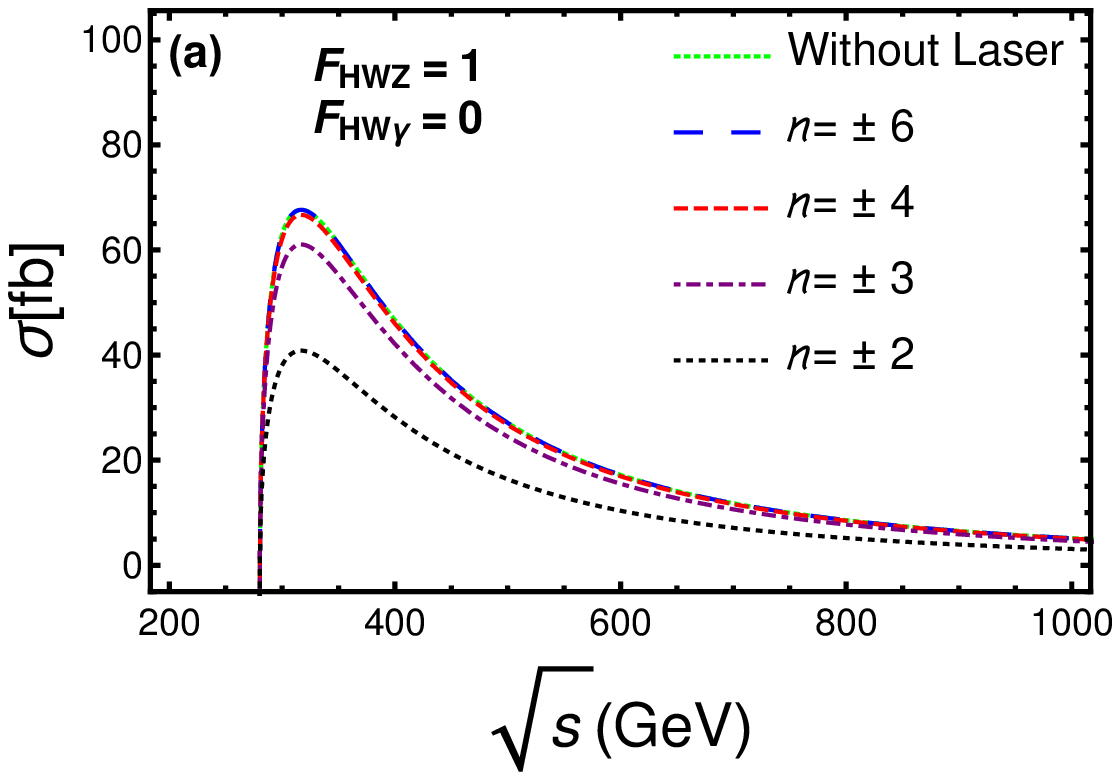}
    \label{fig2}
  \end{minipage}
  \hspace*{0.5cm}
  \begin{minipage}[t]{0.47\textwidth}
  \centering
    \includegraphics[width=\textwidth]{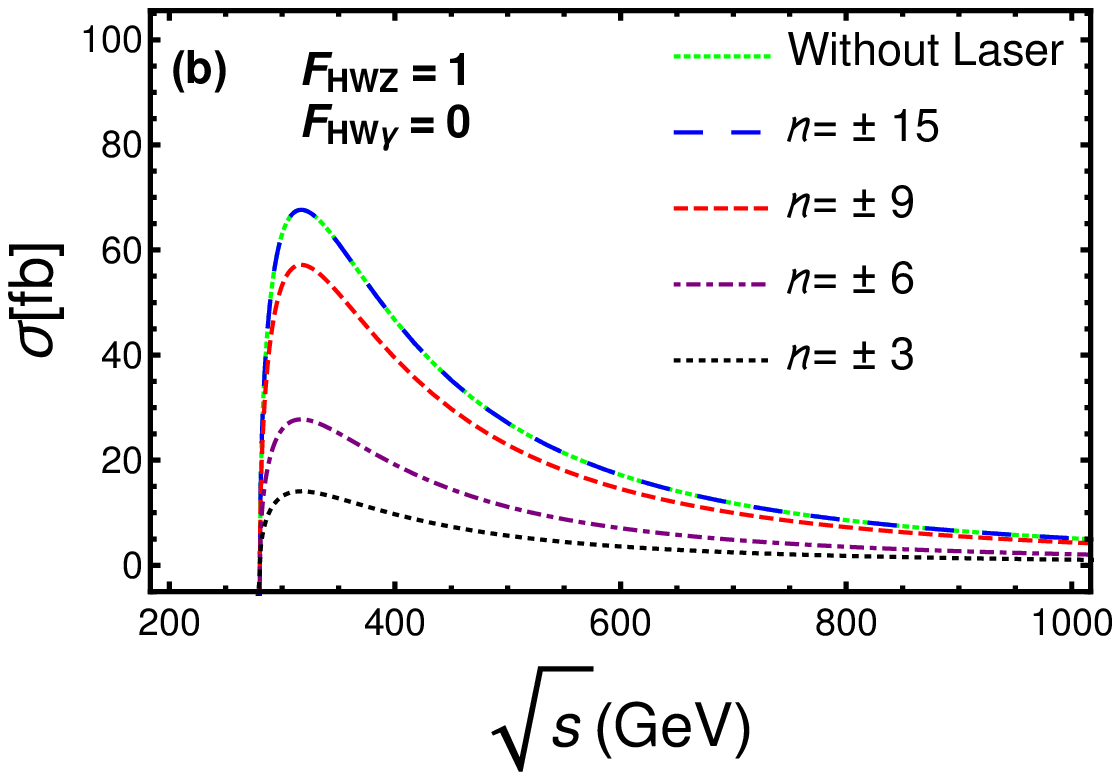}
    \label{fig3}
  \end{minipage}
   \begin{minipage}[t]{0.478\textwidth}
  \centering
    \includegraphics[width=\textwidth]{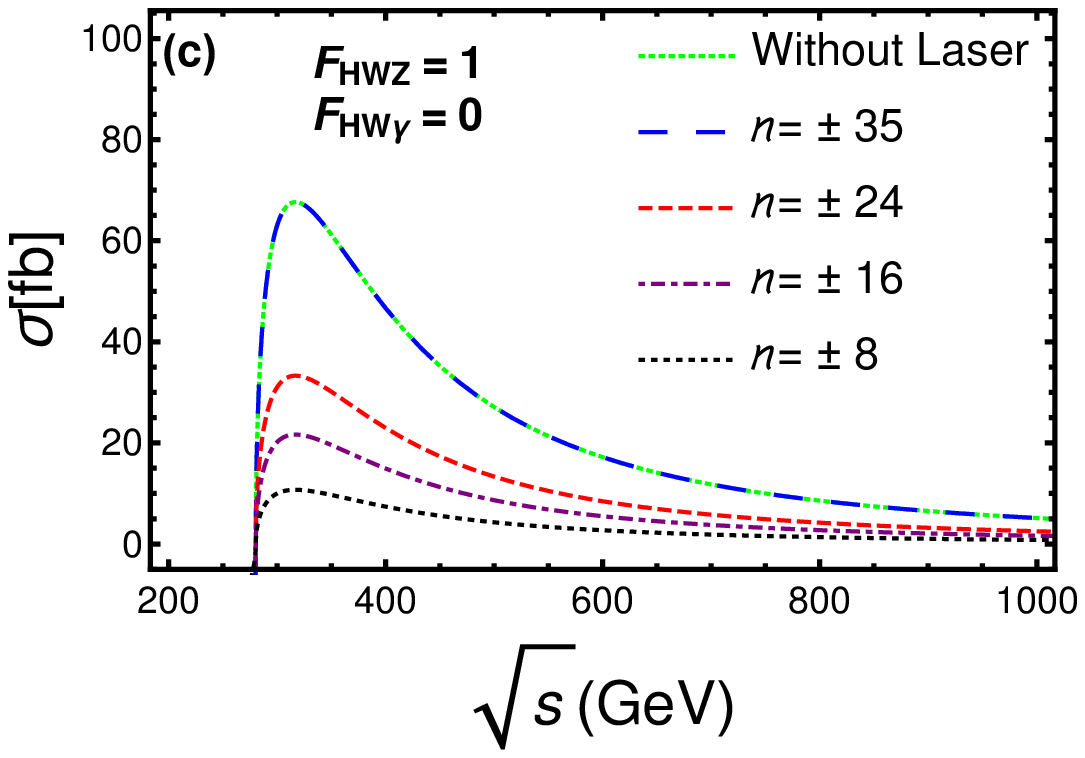}
    \label{fig4}
  \end{minipage}
  \hspace*{0.5cm}
  \begin{minipage}[t]{0.47\textwidth}
  \centering
    \includegraphics[width=\textwidth]{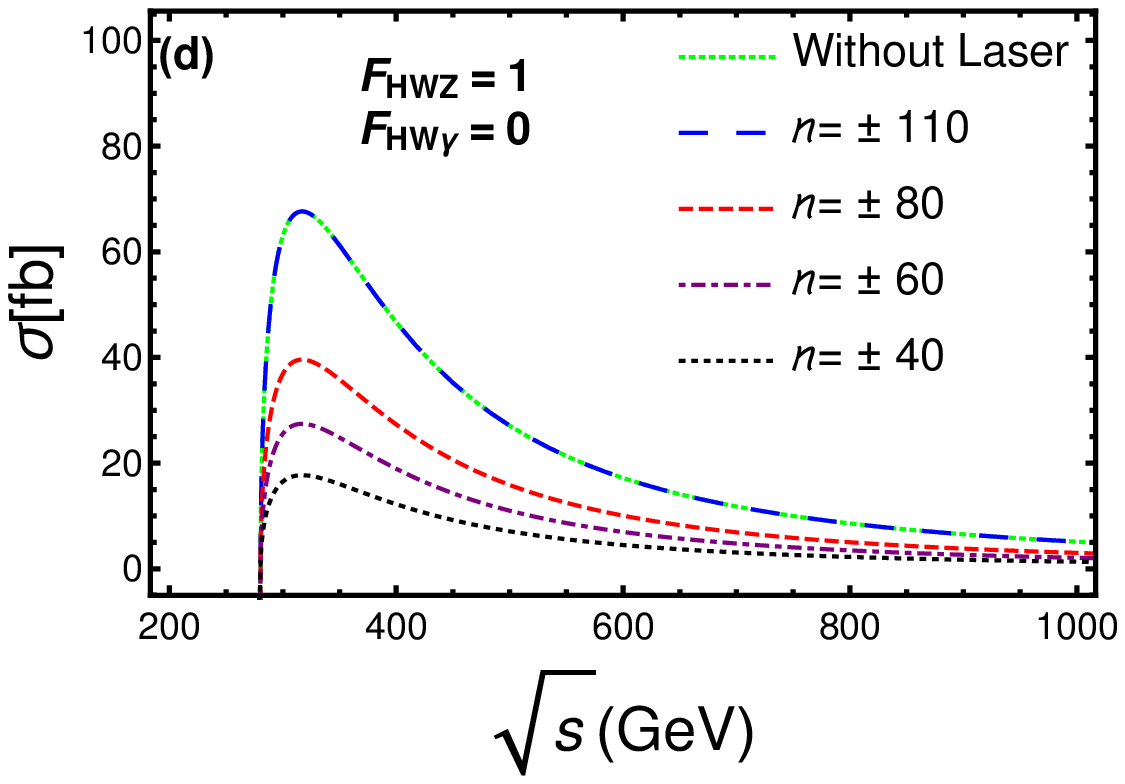}
    \label{fig5}
  \end{minipage}
   \caption{Dependence of the laser-assisted total cross section of the process ${e}^{+}{e}^{-}\rightarrow H^{\pm}W^{\mp}$ on $\sqrt{s}$ in the case where $F_{HWZ}=1$ and $ F_{HW\gamma}=0$ for different exchanged photons number and by taking the charged Higgs mass as $M_{H^{\pm}}=200\,GeV$. The laser field strength and its frequency are chosen as: (a): $\varepsilon_{0}=10^{5}\,V.cm^{-1}$ , $\,\omega=2\, eV$ ; (b): $\varepsilon_{0}=10^{5}\,V.cm^{-1}$ , $\,\omega=1.17\, eV$ ; (c): $\varepsilon_{0}=10^{6}\,V.cm^{-1}$ , $\,\omega=2\, eV$ ; (d): $\varepsilon_{0}=10^{6}\,V.cm^{-1}$ , $\,\omega=1.17 \,eV$.}
   \label{fig3}
\end{figure}
To answer this question, let's begin by analyzing the dependence of the total cross section on the mass of the charged Higgs. Figure \ref{fig4} shows the variation of the total cross section as a function of $M_{H^{\pm}}$ for three known laser frequencies (left panel) and for different laser field strengths (right panel) by summing over $n$ from $-4$ to $+4$.
According to this figure, we find that the total cross section of ${e}^{+}{e}^{-}\rightarrow H^{\pm}W^{\mp}$ reaches up to $85 - 86\, fb$. This maximal value of the cross section is reached in the range of intermediate masses of $H^{\pm}$ ( $M_{H^{\pm}}\sim 150 - 176\,GeV$) for $\omega=2\, eV$ and $\varepsilon_{0}=10^{5}\, V.cm^{-1}$. Therefore, the cross section decreases by increasing the mass of the charged Higgs. This effect is due to the fact that the cross-section is inversely proportional to the mass $M_{H^{\pm}}$. Another important observation which should be discussed here is that the order of magnitude of the cross section is dropping rapidly as long as the laser frequency decreases (left panel). 
For instance, in the low masses region ($M_{H^{\pm}}\,<\,160 \,GeV$), the total cross section reaches up to $85\,fb$ for $\omega=2\, eV$, while it remains under $24\,fb$ for $\omega=1.17\, eV$. Moreover, for $\omega=0.117\, eV$, the total cross section's order is very close to zero. 
This is due to the presence of a strong correlation between the number of photons needed to fulfill the well-known sum rule and the frequency $\omega$. In general, this number increases so far as the frequency of the laser field decreases. 
In the right panel of figure \ref{fig4}, we can see that the order of magnitude of the cross section decreases by increasing the laser field strength, specifically, as long as $\varepsilon_{0}\,\geq\,10^{5}\,V.cm^{-1}$. 
However, we find that the laser-assisted cross section is equal to its corresponding laser-free cross section for $\varepsilon_{0}\,\leq\,10^{4}\,V.cm^{-1}$. Therefore, the electromagnetic field does not affect the integrated laser-assisted cross section at low strengths. 
To confirm these results, we have made the same analysis for the same laser field strengths and frequencies. 
In figure \ref{fig5}, we have plotted the variation of laser-assisted total cross section versus the centre of mass energy for different strengths of the electromagnetic field (left panel) and different frequencies (right panel). As this figure shows, the peak does not change for all laser field's parameters. Consequently, it confirms the validity of our previous results.
\begin{figure}[h!]
\centering
  \begin{minipage}[t]{0.478\textwidth}
  \centering
    \includegraphics[width=\textwidth]{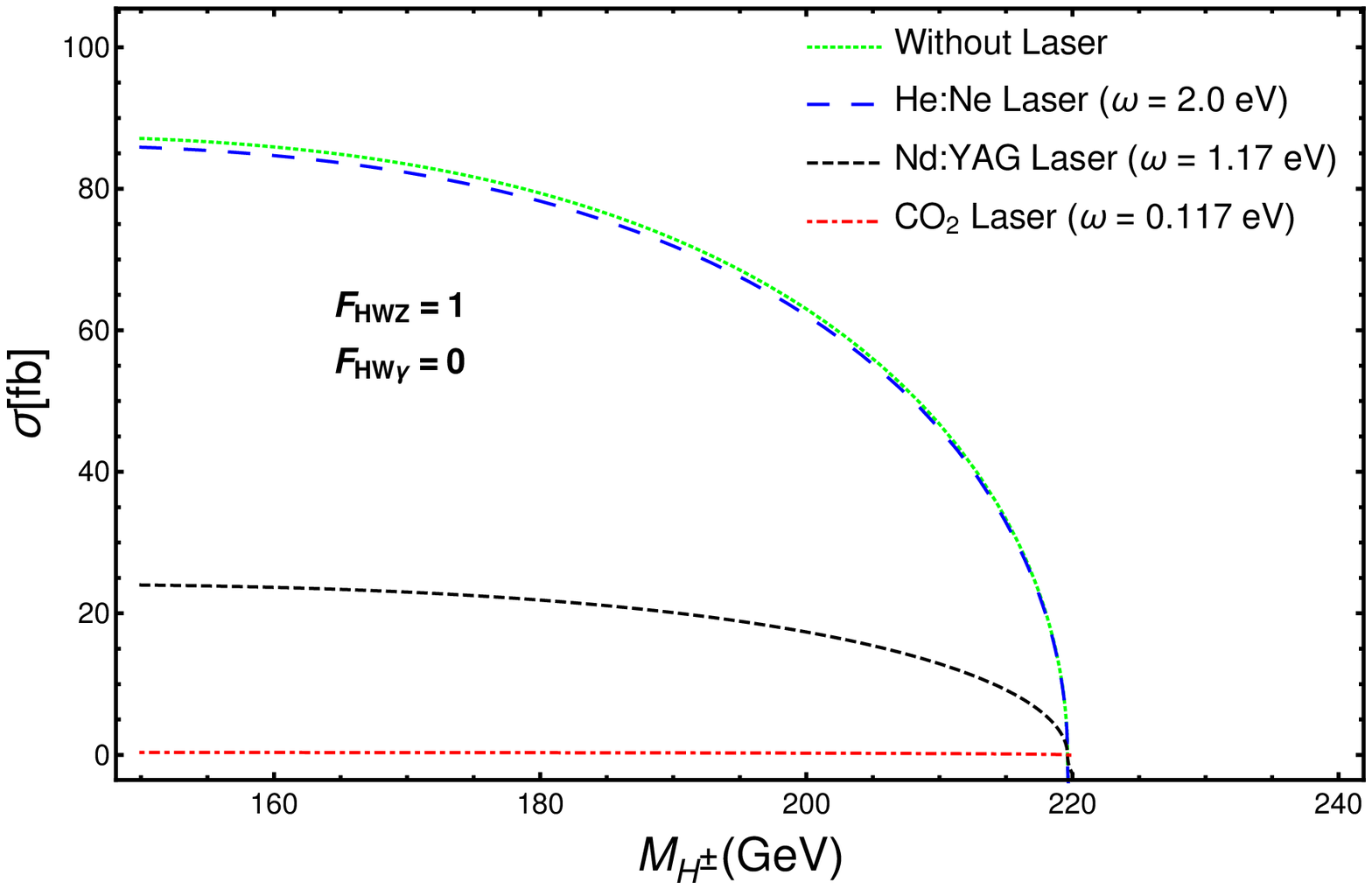}
    \label{fig2}
  \end{minipage}
  \hspace*{0.5cm}
  \begin{minipage}[t]{0.47\textwidth}
  \centering
    \includegraphics[width=\textwidth]{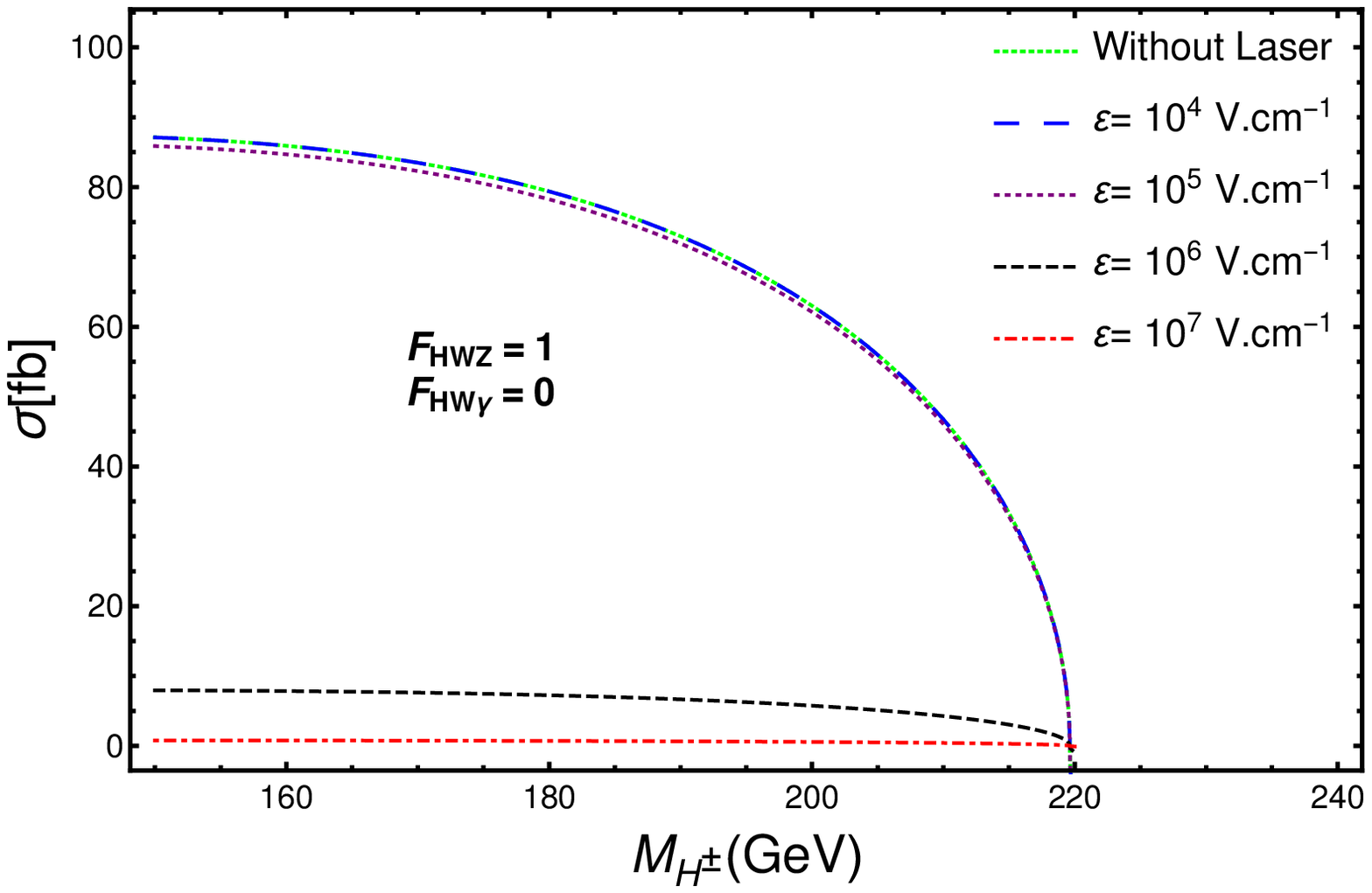}
    \label{fig3}
  \end{minipage}
   \caption{Laser-assisted total cross section of the process ${e}^{+}{e}^{-}\rightarrow H^{\pm}W^{\mp}$ as a function of the charged Higgs-boson for different laser field strengths and frequencies in the case where $F_{HWZ}=1$ and $ F_{HW\gamma}=0$. The $\sqrt{s}$ and $n$ are taken as $300\,GeV$ and $\pm 4$, respectively. (left panel) $\varepsilon_{0}=10^{5}\,V.cm^{-1}$; (right panel) $\omega=2\, eV$.}
        \label{fig4}
\end{figure}

\begin{figure}[h!]
\centering
  \begin{minipage}[t]{0.478\textwidth}
  \centering
    \includegraphics[width=\textwidth]{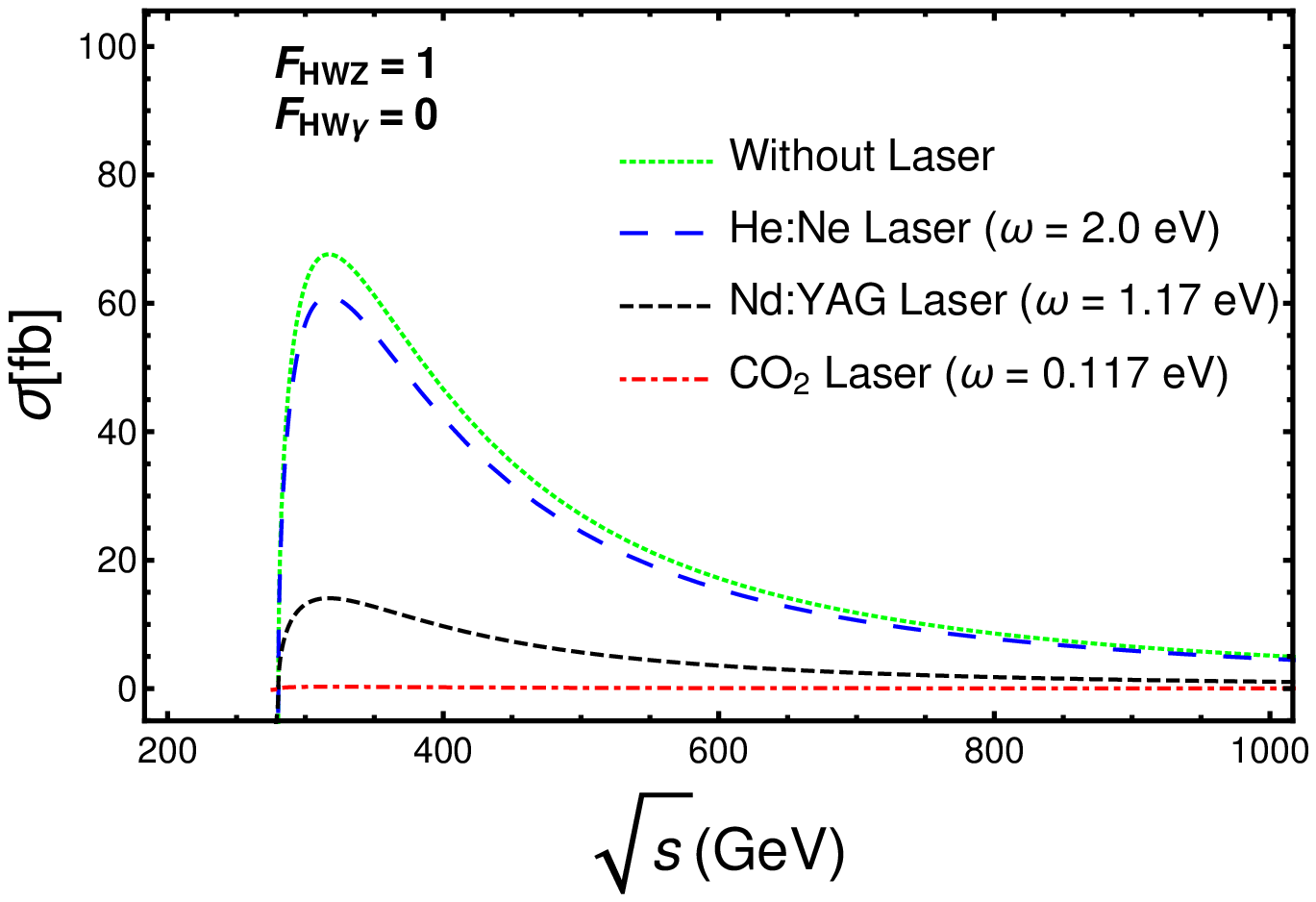}
    \label{fig2}
  \end{minipage}
  \hspace*{0.5cm}
  \begin{minipage}[t]{0.47\textwidth}
  \centering
    \includegraphics[width=\textwidth]{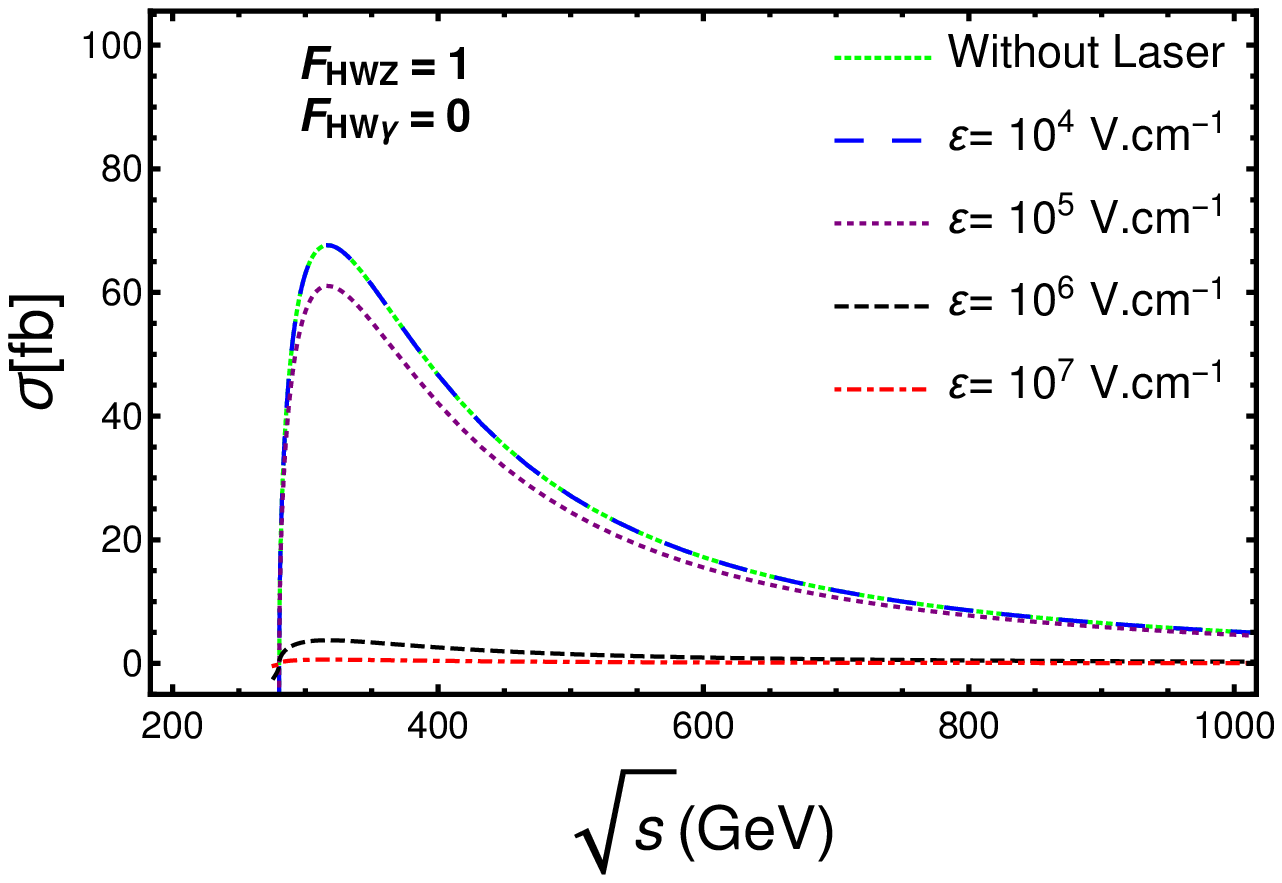}
    \label{fig3}
  \end{minipage}
   \caption{Laser-assisted total cross section of the process ${e}^{+}{e}^{-}\rightarrow H^{\pm}W^{\mp}$ as a function of the centre of mass energy for different laser field strengths and frequencies in the case where $F_{HWZ}=1$ and $ F_{HW\gamma}=0$. The charged Higgs mass and the number of exchanged photons are chosen as $M_{H^{\pm}}=200\,GeV$ and $n=\pm 4$, respectively. (left panel) $\varepsilon_{0}=10^{5}\,V.cm^{-1}$; (right panel) $\omega=2\, eV$.}
        \label{fig5}
\end{figure}

\begin{figure}[h!]
\centering
  \begin{minipage}[t]{0.478\textwidth}
  \centering
    \includegraphics[width=\textwidth]{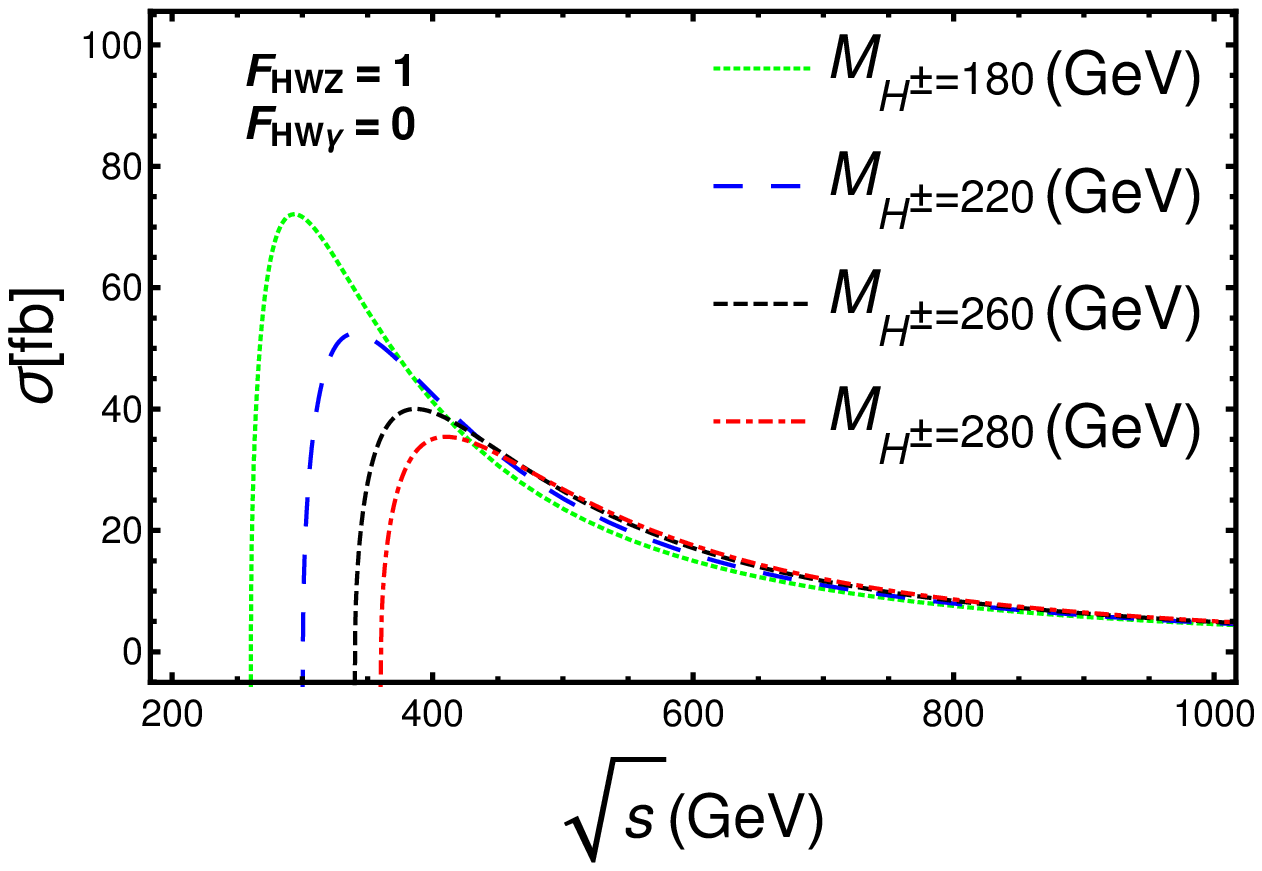}
    \label{fig2}
  \end{minipage}
  \hspace*{0.5cm}
  \begin{minipage}[t]{0.47\textwidth}
  \centering
    \includegraphics[width=\textwidth]{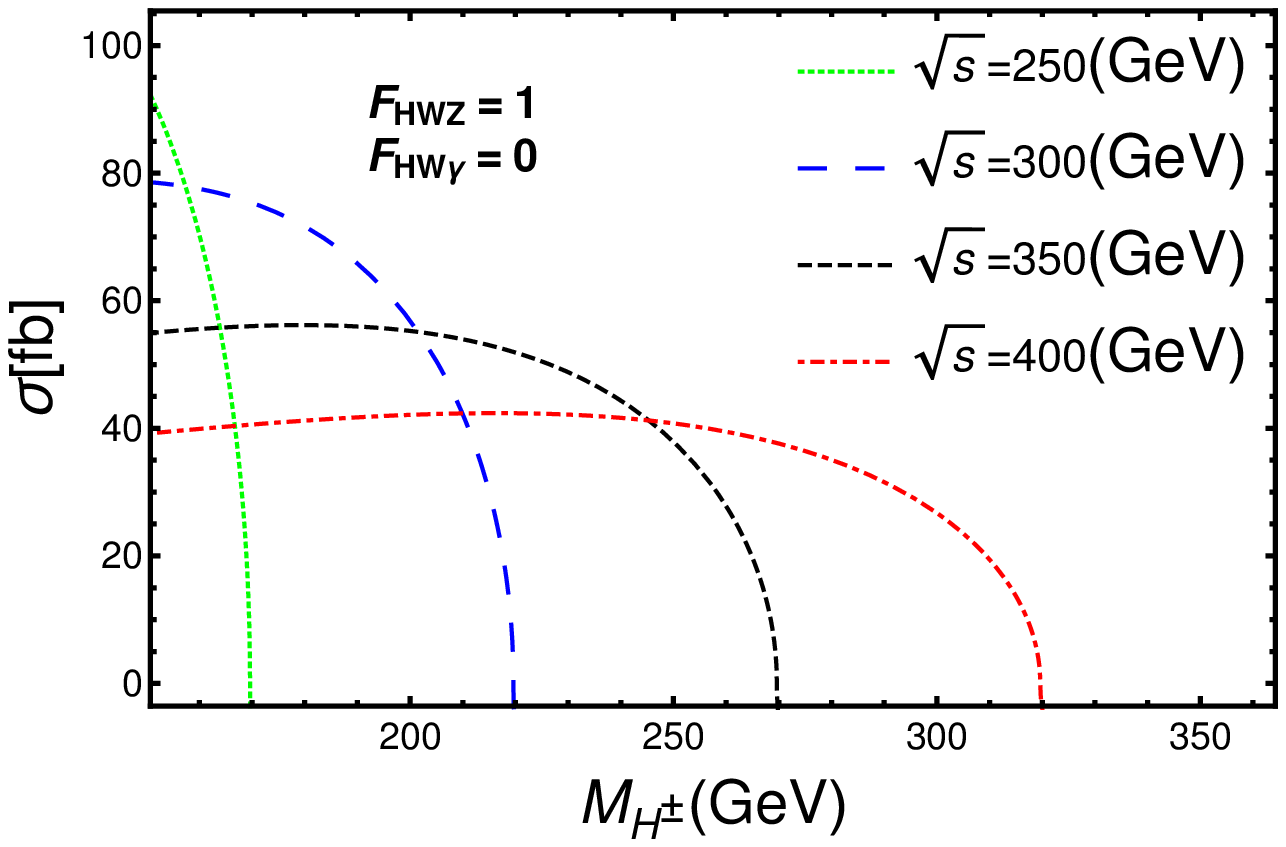}
    \label{fig3}
  \end{minipage}
   \caption{Laser-assisted total cross section of the process ${e}^{+}{e}^{-}\rightarrow H^{\pm}W^{\mp}$ for some typical values of $M_{H^{\pm}}$ (left panel) and $\sqrt{s}$ (right panel) in the case where $F_{HWZ}=1$ and $ F_{HW\gamma}=0$. The number of exchanged photons, the laser field strength and its
frequency are chosen such that: $n=\pm 4$, $\varepsilon_{0}=10^{5}\,V.cm^{-1}$ and $\omega=2\, eV$, respectively.}
        \label{fig6}
\end{figure}
As a last result, we illustrate in figure \ref{fig6} the dependence of the laser-assisted total cross section on the centre of mass energy and the charged Higgs mass for different frequencies and for some typical values of $M_{H^{\pm}}$ (left panel) and centre of mass energies (right panel). The laser field parameters are taken such as $\varepsilon_{0}=10^{5}\, V.cm^{-1}$ and $\omega=2\, eV$. As we can see from the left panel of figure \ref{fig6}, at low values of $\sqrt{s}$ ($\sqrt{s}<430\, GeV$), the $M_{H^{\pm}}=180\, GeV$ scenario dominates, followed by $M_{H^{\pm}}=220\, GeV$ and then $M_{H^{\pm}}=260\, GeV$. In addition, there is a peak at $\sqrt{s}\simeq 290\,GeV$ for $M_{H^{\pm}}=180\, GeV$ while it is located at $\sqrt{s}\simeq 335\,GeV$, $\sqrt{s}\simeq 385\,GeV$ and $\sqrt{s}\simeq 420\,GeV$ for $M_{H^{\pm}}=220\, GeV$, $M_{H^{\pm}}=260\, GeV$ and $M_{H^{\pm}}=280\, GeV$, respectively. Whereas, at large values of $\sqrt{s}$ ($\sqrt{s}>430\, GeV$), we can see that the curves which represent the $\sqrt{s}$ dependence of the cross section are very close to each other for all values of $M_{H^{\pm}}$. In figure \ref{fig6} (right panel), we can see that the order of magnitude of the cross section depends on the centre of mass energy. For example, its maximum values are $\sigma\simeq90fb$, $\sigma\simeq78fb$, $\sigma\simeq57fb$ and $\sigma\simeq43fb$ for $\sqrt{s}=250\,GeV$, $\sqrt{s}=300\,GeV$, $\sqrt{s}=350\,GeV$ and $\sqrt{s}=400\,GeV$, respectively. By comparing these plots, we can deduce that the cross section drops rapidly for $\sqrt{s}=250\,GeV$, while it decreases slowly for high centre of mass energies (threshold effect). 
\section{Conclusion}\label{sec:Conclusion}
In summary, we have investigated the process which acts as a source of charged Higgs boson associated with W-boson at tree level in the presence of a circularly polarized electromagnetic field. By theoretically analyzing this process, we have presented the dependence of the partial total cross section on the number of exchanged photons, and we have shown that the maximum number of photons that can be exchanged depends only on the laser field parameters. Then, we have found that there is a strong correlation between the order of magnitude and the laser field parameters such as the number of exchanged photons, the laser field strength and its frequency. More precisely, this order of magnitude of the total laser-assisted cross section decreases as long as the laser field strength increases or by decreasing its frequency. However, radiations with low intensities have no effect on the cross section. Furthermore, we have shown that, for a given laser field strength and frequency, the laser-assisted total cross section falls down rapidly for $\sqrt{s}=250\, GeV$ as compared to high values of centre of mass energy.


\section*{References}
\begin{thebibliography}{99}
\bibitem{1}P. Francken, C. J. Joachain. Electron–atomic-hydrogen elastic collisions in the presence of a laser field. Phys. Rev. A \textbf{35}, 1590 (1987). \url{https://doi.org/10.1103/PhysRevA.35.1590}

\bibitem{2}Y. Attaourti and B. Manaut, Comment on “Mott scattering in strong laser fields”
. Phys. Rev. A \textbf{68}, 067401. \url{https://doi.org/10.1103/PhysRevA.68.067401};
Y. Attaourti, B. Manaut, and S. Taj, Mott scattering in an elliptically polarized laser field . Phys. Rev. A \textbf{70}, 023404. \url{https://doi.org/10.1103/PhysRevA.70.023404};
Y. Attaourti, B. Manaut, and A. Makhoute, Relativistic electronic dressing in laser-assisted electron-hydrogen elastic collisions . Phys. Rev. A \textbf{69}, 063407. \url{https://doi.org/10.1103/PhysRevA.69.063407};

\bibitem{3}Y. Attaourti, S. Taj, and B. Manaut, Semirelativistic model for ionization of atomic hydrogen by electron impact, Phys. Rev. A \textbf{71}, 062705. \url{https://doi.org/10.1103/PhysRevA.71.062705};
B. Manaut, S. Taj, and Y. Attaourti, Mott scattering of polarized electrons in a strong laser field, Phys. Rev. A \textbf{71}, 043401 . \url{https://doi.org/10.1103/PhysRevA.71.043401}.

\bibitem{4} M.Jakha, S.Mouslih, S.Taj, Y.Attaourti, B.Manaut, Influence of intense laser fields on measurable quantities in $W^{-}$-boson decay, \url{https://doi.org/10.1016/j.cjph.2021.09.011}.

\bibitem{5} M.Baouahi, M Ouali, M.Jakha, S.Mouslih, Y.Attaourti, B.Manaut1, S.Taj, Laser-assisted kaon decay and CPT symmetry violation,
 Laser Phys. Lett. \textbf{18}, 106001 (2021). \url{https://doi.org/10.1088/1612-202X/ac1e86}.

\bibitem{6}S.W. Bahk, et al, Generation and characterization of the highest laser intensities ($10^{22}\, W/cm^{2}$),  Optics Lett.\textbf{29} (2004) 2837, \url{https://doi.org/10.1364/OL.29.002837}

\bibitem{7} M. Ouhammou, M. Ouali, S. Taj, B. Manaut, Laser-assisted neutral Higgs-boson pair production in Inert Higgs Doublet Model (IHDM), 2021, Chin. J. Phys, \url{https://doi.org/10.1016/j.cjph.2021.09.012}.

\bibitem{8} M. Ouali, M. Ouhammou, S. Taj, R. Benbrik, B. Manaut, Laser-assisted charged Higgs pair production in Inert Higgs Doublet Model (IHDM), Phys.Lett.B \textbf{823} (2021)  136761
\url{https://doi.org/10.1016/j.physletb.2021.136761}.

\bibitem{9}M. Ouhammou, M. Ouali, S. Taj, and B. Manaut, Higgs-strahlung boson production in the presence of a circularly polarized laser field, Laser Phys. Lett. \textbf{18}, 076002 (2021). \url{https://doi.org/10.1088/1612-202X/ac0919}.


\bibitem{ouali} M. Ouali. M. Ouhammou. Y. Mekaoui. S. Taj. B. Manaut, $Z-$boson production via the weak process $\e^{+}e^{-}\rightarrow \mu^{+}\mu^{-}$ in the presence of a circularly polarized laser field, \emph{Chinese Journal of Physics}, (2021) \url{https://doi.org/10.1016/j.cjph.2021.10.007}; Y. Mekaoui, M. Jakha, S. Mouslih, B. Manaut, R. Benbrik, S. Taj, Relativistic elastic scattering of an electron by a muon in the field of a circularly polarized electromagnetic wave, 
\url{arXiv:2110.06695}

\bibitem{10} S. Mouslih, M. Jakha, S. Taj, B. Manaut and E. Siher, Laser-assisted pion decay, Phys. Rev. D \textbf{102} (2020) 073006,
\url{https://doi.org/10.1103/PhysRevD.102.073006}.

\bibitem{12}
G.~Aad \textit{et al.}, [ATLAS], Observation of a new particle in the search for the Standard Model Higgs boson with the ATLAS detector at the LHC,
Phys. Lett. B \textbf{716} (2012) 1, 
\url{https://doi.org/10.1016/j.physletb.2012.08.020}

\bibitem{13}S.~Chatrchyan \textit{et al.} [CMS], Observation of a new boson at a mass of 125 GeV with the CMS experiment at the LHC,
Phys. Lett. B \textbf{716} (2012) 30,
\url{https://doi.org/10.1016/j.physletb.2012.08.021}

\bibitem{14}M. Hashemi and Ijaz Ahmed, Observability of triple or double charged Higgs production in two Higgs doublet model type II at an $e^{+}e^{-}$ linear collider Int.J.Mod.Phys.A \textbf{30} (2015)
\url{https://doi.org/10.1142/S0217751X15500220}

\bibitem{15}A. Arhrib, R. Benbrik, M. Krab, B. Manaut, S. Moretti, Yan Wang, Qi-Shu Yan, New discovery modes for a light charged Higgs boson at the LHC, JHEP \textbf{10} (2021) 073
\url{https://doi.org/10.1007/JHEP10(2021)073}

\bibitem{J.Ouaali}J. Ou aali, B. Manaut, L. Rahili, S. Semlali, Naturalness implications within the two-real-scalar-singlet beyond the SM, Eur.Phys.J.C \textbf{81}, 1045 (2021).
\url{https://doi.org/10.1140/epjc/s10052-021-09839-6}

\bibitem{Tania}T. Robens, T. Stefaniak, J. Wittbrodt, Two-real-scalar-singlet extension of the SM: LHC phenomenology and benchmark scenarios, Eur.Phys.J.C \textbf{80} (2020)
\url{https://doi.org/10.1140/epjc/s10052-020-7655-x}

\bibitem{19}H.Abouabid, A.Arhrib, R.Benbrik, J.El Falaki, B.Gong, W.Xie, Q.Yan, One-loop radiative corrections to $\e^{+}e^{-}\rightarrow Zh^{0}/H^{0}A^{0}$ in the Inert Higgs Doublet Model, JHEP \textbf{05} (2021) 100 \url{https://doi.org/10.1007/JHEP05(2021)100} 

\bibitem{20}I.Ahmed, F.Khaliq, T.Khurshid 
	arXiv:2003.08193 [hep-ph]
	
\bibitem{21}L. Rahili, A. Arhrib, R. Benbrik, Associated production of SM Higgs with a photon in type-II seesaw models at the ILC, Eur.Phys.J.C \textbf{79} (2019)	
\url{https://doi.org/10.1140/epjc/s10052-019-7471-3}

\bibitem{22}S.Kanemura, Possible enhancement of the $\e^{+}e^{-}\rightarrow H^{\pm}W^{\mp}$ cross section in the two-Higgs-doublet model, Eur.Phys.J.C \textbf{17} (2000) 473-486
\url{https://doi.org/10.1007/s100520000480}

\bibitem{23}S.Kanemura, K.Yagyu, K.Yanase, Testing Higgs models via the $H^{\pm}W^{\mp}Z$ vertex by a recoil method at the International Linear Collider  Phys.Rev.D \textbf{83} (2011) 075018
\url{https://doi.org/10.1103/PhysRevD.83.075018}

\bibitem{eff1}M.C. Peyran\`ere, H.E, Haber, P. Irulegui, $H^{\pm}\rightarrow W^{\pm}\gamma$ and $H^{\pm}\rightarrow W^{\pm}Z$ in two-Higgs-doublet models: Large-fermion-mass limit, Phys.Rev.D \textbf{44}, 191 (1991)
\url{https://doi.org/10.1103/PhysRevD.44.191}

\bibitem{eff2} J.L.Diaz-Cruz, J.H. S\'anchez, J.J. Toscano, An effective Lagrangian description of charged Higgs decays $H^{\pm}\rightarrow W^{\pm}\gamma, W^{\pm}Z$ and $W^{\pm}h^{0}$, Phys.Lett.B \textbf{512} (2001) 339–348
\url{https://doi.org/10.1016/S0370-2693(01)00703-1}

\bibitem{25}E.Asakawa, S.Kanemura, The $H^{\pm}W^{\mp}Z$ vertex and single charged Higgs boson production via $WZ$ fusion at the Large Hadron Collider,  Phys.Lett.B \textbf{626} (2005) 111-119 
\url{https://doi.org/10.1016/j.physletb.2005.08.091}

\bibitem{28}W. Greiner and B. Mueller, Gauge Theory of Weak Interactions, 3rd ed. (Springer, Berlin, 2000).

\bibitem{Feyncalc} R. Mertig, M. Bohm, and A. Denner, Feyn Calc- Computer-algebraic calculation of Feynman amplitudes. Comput. Phys. Commun. \textbf{64}, (1991) 345-359. \url{http://dx.doi.org/10.1016/0010-4655(91)90130-D}; V. Shtabovenko, R. Mertig, and F. Orellana, New developments in FeynCalc 9.0, Comput. Phys. Commun. \textbf{207}, (2016) 432-444. \url{http://dx.doi.org/10.1016/j.cpc.2016.06.008}.

\bibitem{29}P.A. Zyla et al. (Particle Data Group), Review of Particle Physics, 2020, Prog. Theor. Exp. Phys. \textbf{2020}, 083C01. \url{https://doi.org/10.1093/ptep/ptaa104}.

\bibitem{30}Q. Yang, R.Y. Zhang, M.M.Long S.M. Wang W.G. Ma J.W. Zhu Y. Jiang, QCD Corrections to $\e^{+}e^{-}\rightarrow H^{\pm}W^{\mp}$ in Type-I THDM at Electron Positron Colliders, 
 Chin.Phys.C \textbf{44} (2020) 9, 9
\url{arXiv:2005.11010 [hep-ph]}

\bibitem{31}S.H. Zhu, hep-ph/9901221.

\bibitem{32}S. Heinemeyer, C. Schappacher, Charged Higgs Boson production at $e^{+}e^{-}$ colliders in the complex MSSM: a full one-loop analysis, 
Eur.Phys.J.C \textbf{76} (2016) 10, 535
\url{https://doi.org/10.1140/epjc/s10052-016-4383-3}

\bibitem{33}F. V. Bunkin and M. V. Fedorov, Sov. Phys. JETP \textbf{22} (1966) 844 ; N. M. Kroll and K. M. Watson, Phys. Rev. A \textbf{8} (1973) 804.

\end{thebibliography}
\end{document}